\documentclass[fleqn]{aa}
\usepackage[varg]{txfonts}
\usepackage{graphicx, subfigure}
\usepackage{epstopdf}
\usepackage{natbib}
\usepackage{verbatim}
\usepackage{txfonts,epsfig,url,twoopt}
\usepackage[breaklinks=true]{hyperref}
\hypersetup{colorlinks=true,citecolor=blue}
\bibpunct{(}{)}{;}{a}{}{,} 
\usepackage{xcolor} 
\usepackage{cleveref}
\usepackage{nicefrac}
\usepackage{booktabs,multirow}
\usepackage{siunitx}

\definecolor{sigmacolor}{HTML}{800080}
\definecolor{cavitycolor}{HTML}{007F00}
\newcommand{\INNERBAU}{Disc inner boundary\xspace}
\newcommand{\innerbou}{disc inner boundary\xspace}
\newcommand{\ib}{IB\xspace}
\newcommand{\DEADZONE}{Dead-zone inner edge\xspace}
\newcommand{\deadzone}{dead-zone inner edge\xspace}
\newcommand{\dz}{DZ\xspace}

\begin{document}
	\title{Pushing planets into an inner cavity by a resonant chain}	

	\titlerunning{Trapping planets at a disc inner edge}
	\authorrunning{S.~Ataiee \& W.~Kley}
	
	\author{S.~Ataiee \inst{\ref{inst1}, \ref{inst2}} \thanks{sareh.ataiee@uni-tuebingen.de}
		\and 
		W.~Kley \inst{\ref{inst1}}
	}
	\institute{Institut f\"ur Astronomie \& Astrophysik, Universit\"at T\"ubingen, Auf der Morgenstelle 10, 72076 T\"ubingen, Germany \label{inst1}
				\and
				Department of Physics, Faculty of Sciences, Ferdowsi University of Mashhad, Mashhad, 91775-1436, Iran \label{inst2}
	}
	
	\abstract
{The orbital distribution of exoplanets indicates an accumulation of  super-Earth sized planets close to their host stars in compact systems. When an inward disc-driven migration scenario is assumed for their formation, these planets could have been stopped and  might have been parked at an inner edge of the disc, or be pushed through the inner disc cavity by a resonant chain. This topic has not been properly and extensively studied.}
{Using numerical simulations, we investigate the possibility that the inner planets in a resonant chain can be pushed into the disc inner cavity by outer planets.}
{We  performed hydrodynamical and N-body simulations of planetary systems embedded in their nascent disc. The inner edge of the disc was represented in two different ways, resembling either a dead zone inner edge (DZ) or a disc inner boundary (IB). The main difference lies in the steepness of the surface density profile. The innermost planet always has a mass of $10\, \rm M_{Earth}$, with additional outer planets of equal or higher mass.}
{A steeper profile is able to stop a chain of planets more efficiently than a shallower profile. 
The final configurations in our DZ models are usually tighter than in their IB counterparts, and therefore more prone to instability. 
We derive analytical expressions for the stopping conditions based on power equilibrium, and show that the final eccentricities result from torque equilibrium. 
For planets in thinner discs, we found, for the first time, clear signs for over-stable librations in the hydrodynamical simulations, leading to very compact systems.
We also found that the popular N-body simulations may overestimate the number of planets in the disc inner cavity.
}
{}
	
	
	\maketitle
	\section{Introduction}
	\label{sec:into}
		The notable number of short-period super-Earths is one of the puzzles introduced by the exoplanet discovery surveys \cite[e.g.][]{2015ARA&A..53..409W,2015ApJ...807...45D}. To answer the question of how these planets could accumulate at sub-au proximity to their host stars, several scenarios have been proposed, such as a combination of planet trapping at the magnetospheric truncation radius of the disc and star-planet tidal interaction \citep{2017ApJ...842...40L}, planet-planet interaction aided by the eccentricity damping of the gas disc \citep{2008ApJ...686..580C,2009ApJ...699..824O}, planet trapping at the disc inner edge \citep{2011A&A...528A...2B,2018MNRAS.473.5267M}, and pushing of the inner planets in a resonant chain by the outer planets in the system \citep{2016MNRAS.457.2480C,2019MNRAS.486.3874C}. The common core idea in all of these scenarios is the existence of a disc inner edge,  at which the planets can be trapped. This trapping can either lead to the formation of a resonant chain that might become unstable, push the inner planets into the disc inner cavity, or simply stall the migration a single planet.
		
		Soon after the discovery of the first exoplanet, 51 Peg\,b, it was suggested that this short-period planet has been stopped by the inner edge of its natal disc \citep{1996Natur.380..606L}. By the means of two-dimensional (2D) hydrodynamical simulations, \cite{2006ApJ...642..478M} showed that the migration of a low-mass planet can be halted at a steep surface density transition by the change in the Lindblad torque and enhancement of the positive co-rotation torque. In order to sustain a transition in the surface density, the authors increased the viscosity inside the inner disc in their models. This viscosity transition can itself modify the co-rotation torque. Without assuming a transition in the viscosity, \cite{2019MNRAS.485.2666R} supported the trapping of low-mass planet at the disc inner edge by three-dimensional (3D) hydrodynamical simulations. In their study, the disc inner edge is modelled as a region with a transition in density and temperature such that the pressure equilibrium is sustained. Based on this, the trapping of a single low-mass planet at the inner edge of the disc seems to be a robust phenomenon. However, it is still unclear whether an inner edge could also stop the migration of multiple planets.
		
		Trapping of multi-planet systems at the disc inner edge is often observed in N-body simulations \citep[e.g][]{2017MNRAS.470.1750I,2018MNRAS.479L..81R}. In these simulations, the migration and eccentricity damping of the planets are modelled using the formulae that are obtained from the results of hydrodynamical simulations for a single planet in a power-law disc. It has not been investigated whether these formulae are applicable close to the inner edge of a disc, where a sharp gradient in the surface density exists. On the other hand, the trapping of multiple planets in studies that performed hydrodynamical simulations has not been vigorously investigated. \cite{2005MNRAS.363..153P} and \cite{2019ApJ...872...72C} inspected the convergent migration of low-mass two-planet systems in a disc with a surface density profile that resembled the disc inner edge. The surface density in the inner part of their disc rose linearly from the inner boundary up to a given distance and then became constant. In the outer part it became a decreasing power-law. The authors did not report whether the inner planet is pushed into the inner region of the disc, except in one figure of \cite{2005MNRAS.363..153P}, in which the planets become very close to the inner boundary. This means that the trapping of multiple planets using hydrodynamical simulation needs to be studied more vigorously.
		
		Trapping of multi-planet systems at the disc inner edge can differ from a single planet in two aspects. Firstly, trapped planets are usually in resonance, which results in the excitation of their eccentricities \citep[e.g][]{2005MNRAS.363..153P}. Eccentricity damping of a single planet and multi-planet system has been investigated in power-law discs \citep{2006A&A...450..833C,2008A&A...482..677C,2010A&A...523A..30B}, but it is not obvious whether a trapped eccentric planet at the disc inner edge behaves similarly. \cite{2010ApJ...721.1184O} used analytical calculations along with orbital integration and found that an inner eccentric planet in a resonant chain that is close to the inner edge of a disc can halt the migration of the whole chain. This type of trapping, called eccentricity trap, arises from different eccentricity damping rates of the inner planet at the apocentre and pericentre. Although they did not consider the change in torque on the inner planet corresponding to the surface density jump, their study indicates a different behaviour of the eccentric planets at the disc edge. Secondly, multiple planets in a resonant chain migrate as one entity. Whether trapping the inner planet can stop the whole chain or if a massive enough system would overcome this and break the trap needs to be probed properly.
		
		The outcome of N-body simulations of multiple planets at the disc inner edge highly depends on what equation of motion is used and how the inner edge is modelled \citep{2018ApJ...864L...8B,2020MNRAS.494.5666I}. One advantage of hydrodynamical simulations is that the migration of a planetary system is directly calculated through the disc-planet and planet-planet gravitational interactions. Therefore the main complexity would be the modelling of the inner edge.
		
		The detailed modelling of the inner edge of a disc, where the material is transferred to the central star, can be very complicated \citep{2002ApJ...578..420R,2005ApJ...634.1214L,2006ApJ...645L..73R,2009A&A...508.1117Z}. However, we can simply define it as the location in which the gas leaves the disc and settles on the star. Near the inner edge of the disc, other traps might also exist, such as a dead-zone edge. \cite{2017ApJ...835..230F} studied the inner region of protoplanetary discs around solar-like stars using radiative magnetohydrodynamical simulations and showed that the silicate sublimation at $\sim 0.08$~au leads to the formation of a density jump at $\sim 0.5$~au. This location is the inner edge of the dead-zone, where the turbulence dramatically changes because the ionization degree changes. This means that two radii can exist close to the inner edge, at which planets can be trapped: (1)~the dead-zone edge, and (2)~the inner boundary. They can both trap planets if the disc surface density and temperature profiles provide a strong enough positive torque on the planet. However, the planets have to overcome the traps to approach the cavity between the disc and the star. 
		
		We used 2D hydrodynamical simulations to study the trapping of a planetary system at the inner edge of a disc and also in a dead-zone. We examined resonant systems, mostly composed of super-Earths, and monitored their migration to determine whether the inner planet is able to stop the chain or is pushed into the cavity. For this purpose, we constructed two types of traps that differed in their steepness and long-term evolution. In Sec.~\ref{sec:modelling} we describe how these two types of traps were modelled, which code was used, and how the disc and planets were set up. We performed simulations with two, three, and more planets with different mass combinations, and also some simulations for a lower disc viscosity and lower aspect ratio. The results are listed in Sec.~\ref{sec:results}. After this, we determined in Sec.~\ref{sec:transwidth} whether the results depend on the width of dead-zone edge. A power and torque analysis is presented in Sec.~\ref{sec:torpow}, where we show that the migration of a resonant system halts at the location at which the total power of the system vanishes. In some of our models, we observe an over-stability, which we discuss in Sec.~\ref{sec:overstability}. Finally, our findings are summarised in Sec.~\ref{sec:summ}.

	\section{Modelling the inner edge: Setup and numerics}
	\label{sec:modelling}
		As described in the introduction, it is expected that a protoplanetary disc disrupts near the stellar surface and has an inner edge, where the disc surface density drops significantly. Another location that is also close to the inner edge and has a transition in surface density is the dead-zone edge. Both of them can trap planets if the disc provides a strong enough co-rotation torque. However, the planets have to overcome the trap to get closer to the star. Using 2D hydrodynamical simulations, we studied the migration of planets and their trapping at these two types of density transition. We analysed the final configuration with respect to their resonances and monitored whether the inner planet was pushed into the inner cavity.
			
		Bringing the planets into a resonance, allowing the inner planet to reach the trap, and properly modelling the migration of planets at the disc inner edge is cumbersome and time-consuming. Even when a planetary system is initialised close to a resonance commensurability, it might need thousands of orbits to reach a steady final state. On the other hand, properly resolving the co-rotation torque for a low- or moderate-mass planet needs at least six cells in a half-horseshoe width of the planet \citep{2011MNRAS.410..293P}. A high resolution like this at the distances very close to the star would decrease the time step so dramatically that the simulations become infeasible. To avoid this issue, we shifted the density transition to  about one code length unit. In this way, we were able to properly test whether a resonant migration can push the inner planet out of the trap and into the cavity. 
			
		\subsection{Edge models}
		\label{sec:howtomodel}
		We constructed two types of traps: a \deadzone~(DZ) and a \innerbou~(IB). To facilitate understanding of the results, we plot the DZ and IB simulations in green and purple frames respectively. The same colour code is used in Table~\ref{tab:models} to avoid confusion. These planetary traps are constructed in the ways described below.

		\DEADZONE: We modelled the dead-zone inner edge by decreasing the viscosity over a distance of the order of the disc scale height using the the method of \cite{2006ApJ...642..478M}. In this model the inner part has a higher viscosity and the outer part has a lower value. As a consequence, the inflow velocity is higher inside the cavity and lower outside, which gives rise to a density maximum just outside of the viscosity transition region. The enforced viscosity determines the shape of the density jump. Because the viscosity profile is fixed, it leads to a surface density profile that does not evolve much with time. This fixed surface density profile is the advantage of this model. However, we found that the strong viscosity variation around the location of the planetary trap and the consequent vortex formation complicate the saturation of the co-rotation torque.
			
		\INNERBAU: We assumed that the region between the disc edge and the star is emptied by some mechanism. When the gas reaches the inner edge, it is therefore taken out of the simulations such that an extremely low surface density is created at the inner boundary. The inner disc evolves correspondingly and adopts its profile according to this inner boundary condition. A similar method has been used by \cite{2019MNRAS.485.2666R} to study trapping of a planet in 3D hydro-simulations. The drawback of this model is that this boundary condition causes the surface density profile, which determines both the Lindblad and co-rotation torques, to slowly evolve in time. However, the smooth viscosity can be considered its advantage.

\begin{figure}
				\centerline{\includegraphics[width=\columnwidth]{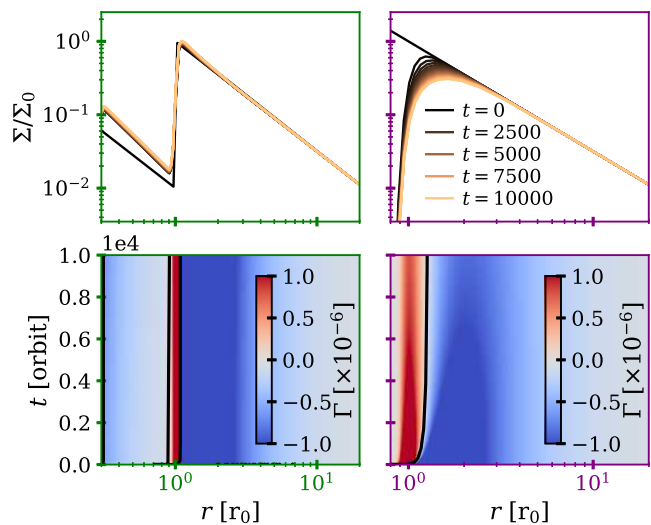}}
				\caption{ Evolution of the disc without any embedded planets for the two types of trap, \dz~(left) and \ib~(right). \textit{Top}: 1D surface density evolution. The colour of the lines represents the time in units of the orbital period at $r=1$. Black denotes the initial profile, and the line colours pale as time proceeds. \textit{Bottom}: Analytical type~I torque acting on a planet with ten Earth masses calculated using the disc profiles. Every horizontal line in these plots shows the specific torque at every distance from the star at a given time. The black contours represent the zero-torque locations. }
				\label{fig:discsetups}
			\end{figure}

		To illustrate the different behaviours of these two traps, we display the evolution of the surface density and the specific torque on an imaginary planet of ten Earth masses in Fig.~\ref{fig:discsetups}. Here and in the subsequent plots, the left panels (green frames) refer to the  DZ and the right panels (purple frames) to the IB model. The top row shows that the DZ model yields a very steep surface density transition that becomes stationary with time. In the bottom row, 2D maps of the specific torque, $\Gamma$, acting on a fictitious planet of ten Earth masses planet at every location in the disc at a specific time during the disc evolution are displayed. The torque was calculated using the formulae in \cite{2010MNRAS.401.1950P,2011MNRAS.410..293P}. The positive density slope at the inner disc edge produces a region in which the torque acting on the planet becomes positive. The black contours indicate where the torque is zero. The actual planetary trap is at the outer edge of the positive torque region. At the trap, the planet feels a negative torque if it is shifted outwards and a positive torque if slightly displaced inwards. In the \dz~ model, there is a sharp transition from the negative to positive torque at the trap, whereas in the \ib~ model, the gradient of the torque is very shallow around the zero-torque location. In other words, if the planet is pushed slightly inwards, it would feel a very strong positive torque in the \dz~ model and is sent back to the zero-torque location, while in the \ib~ model, the planet would feel a slight positive torque that is easily overcome by the push of an outer planet. Based on the torque map of our disc models, we therefore anticipate that a resonant chain of planets would be more successful in pushing a planet out of the planetary trap in the \ib~ than in the \dz~ model. In the next section, we explain the numerical setups of these two models in more detail.
			
		\subsection{Setup and numerics}
		\label{sec:setup}
			We used a modified version of the 2D hydrodynamical code \texttt{FARGO}\footnote{http://fargo.in2p3.fr/-Legacy-archive-}. This code simulates a locally isothermal disc in the cylindrical coordinate. We used a fixed coordinate frame centred on the star and took the indirect terms on the planets and the disc into account.
			
			The aspect ratio of the disc is $h=h_{0}(r/r_{0})^{0.5}$, and $r_{0}=1$ is our length unit. 
The value of $h_{0}$ is 0.05 for the reference model, but all parameter sequences include models with the lower value of 0.03.
Our time unit, called orbit, is defined as one Keplerian orbital time at $r_{0}$. To be initially in viscous equilibrium, we chose $\Sigma = \Sigma_{0}(r/r_{0})^{-1.5}\, \rm[M_{\star}/r_{0}^2]$ with $\Sigma_{0}=2\times 10^{-4}$ except for a few simulations that had twice of this value. $\Sigma_{0}=2\times 10^{-4}$ corresponds to $\sim 1800~\rm g/cm^2$ for a solar-mass star $\rm M_{\star}=1M_{\odot}$ and $r_{0}=1$~au. We note that this is the basic profile in the disc that is modified at the edge at the inner part of the disc for each disc model, as we explain below.
In this setup, the disc inner edge in our models is located near 1\,au, while in real discs, it is approximately ten times smaller.
Our choice was made for numerical reasons, because a smaller radius would require many more time steps due to the shorter dynamical timescale,
not allowing an extensive parameter study. We study locally isothermal discs here, therefore the results should be scalable to smaller radii using the value $h_{0}=0.03$ \citep{2017ApJ...835..230F}.

			For the disc viscosity $\nu$, we used the $\alpha$-viscosity model $\nu=\alpha c_{\rm s} H$, where $c_{\rm s} = H/\Omega$ is the sound speed, with $\Omega$ being the Keplerian angular velocity, and  $H=hr$ is the disc scale height. In most of the simulations, $\alpha = 5\times10^{-4} $ except in few models, in which the effect of a ten times lower viscosity is examined.
			
			In the \dz~ model, the viscosity becomes 100 times higher inside the active zone. The transition occurs over a distance of $\lambda = 1.2H(r_{\rm t})$ at $r_{\rm t}=1$ using the following function \citep[same method as][]{2006ApJ...642..478M}:
			\begin{equation}
				\label{eq:visc}
				\alpha = 5\times 10^{-4} \times 
						\left\{
			   			\begin{array}{ll}
							100 & \quad r < r_{1} \\
							100^{\frac{r_2-r}{r_2-r_1}} & \quad r_1 \leq r \leq r_2 \\
							1 & \quad r > r_{2}
						\end{array}
						\right.
			\end{equation}
			with $r_{1}=r_{\rm t}-\lambda/2$ and $r_{2}=r_{\rm t}+\lambda/2$. 
			For the \ib~ model, $\alpha$ is constant throughout the whole disc.
			
			In the \dz~ models, the disc is meshed into $N_{\rm r} = 929$ logarithmic segments in the radial, and $N_{\rm \phi}=1024$ equal grids in the azimuthal direction. The inner boundary is the outflow and is located at $r_{\rm in}=0.3$. We set our outer boundary far enough at $r_{\rm out}=20$ to avoid any effect of the outer boundary on the migration of outer planets. All quantities were damped to the initial value within $r=18$ to 20 using the method of \cite{2006MNRAS.370..529D}.
			
			In the \ib~ models, the outer boundary is identical to that in the \dz~ model. The inner boundary is located at $r_{\rm in}=0.8$ in order to have the planetary trap around $r=1$. We changed the radial resolution to $N_{\rm r}=712$ but kept the azimuthal grid the same as in the \dz~ model to have a similar resolution in both models. Following the explanation in Sec.~\ref{sec:howtomodel}, we multiplied the surface density within the interval $r\in [0.8-0.9]$ by a factor of 0.999 in each time step until it reached a floor value of $\Sigma_{\rm floor} = 10^{-11}$.  The radial velocity in this region is that of the outflow. This slow decrease in surface density prevents the production of an instability by guaranteeing that the surface density has enough time to be adapted to the new condition. Any other choice of the inner boundary, either in location or damping width, could produce a different profile and consequently a different commensurability from what we present here.
			
			The inner planet was planted at $r=1.2$ in both models. Other planets are located just outside of the 2:1 commensurability of their inner neighbours. The potentials of planets are gradually applied on the disc during the first 50~orbits. This time is long enough for the bump to be established in both models. In some of the simulations, instead of embedding all of the planets from beginning, we added the next planet after the inner pairs retain a resonance. In these models, the tapering time only applies on the first planet. The smoothing length for the potential of the planets is $r_{\rm sm} = \epsilon H(r)$ with $\epsilon = 0.6$ \citep{2012A&A...541A.123M}. To calculate the torque on the planets, we only used the perturbed surface density to avoid any false resonances \citep{2020A&A...635A.204A}. The mass of the inner planet was always $M_{\rm i}=10\, \rm M_{\oplus}$ where $\rm M_{\oplus}=3\times 10^{-6}\, \rm M_{\star}$ is the mass of Earth for a solar-like star. The remaining planets were more massive than or as massive as the inner planet. We used six planets at most (see Table~\ref{tab:models} for the details).

			\begin{table*}
				\centering
				\tiny
				\begin{tabular}{clcccllc}
					\toprule[2pt]
					Number of planets & Name & $h_{0}$ & $\Sigma_{0}$ & $\alpha$ & $r_{\rm p}$ & $M_{\rm p}\,[\rm M_{\oplus}]$ & Result\\
					\midrule[1.5pt]
					\multirow{2}{*}{1} & 1p10 & 0.05 & $2\times 10^{-4}$ & $5\times 10^{-4}$ & 1.2 & 10 & -- \\
					& 1p1Jup & 0.05 & $2\times 10^{-4}$ & $5\times 10^{-4}$ & 1.6 & 333 & --\\
					\midrule
					\multirow{14}{*}{2} & \multirow{2}{*}{2p10} & \multirow{2}{*}{0.05} & \multirow{2}{*}{$2\times 10^{-4}$} & \multirow{2}{*}{$5\times 10^{-4}$} & \multirow{2}{*}{1.2, 1.95} & \multirow{2}{*}{10,10} & \textcolor{cavitycolor}{6:5} \\
					&&&&&&& \textcolor{sigmacolor}{3:2}\\
					\rule{0pt}{2.5ex}
					& \multirow{2}{*}{2p20} & \multirow{2}{*}{0.05} & \multirow{2}{*}{$2\times 10^{-4}$} & \multirow{2}{*}{$5\times 10^{-4}$} & \multirow{2}{*}{1.2, 1.95} & \multirow{2}{*}{10, 20} & \textcolor{cavitycolor}{4:3} \\
					&&&&&&& \textcolor{sigmacolor}{3:2}\\
					\rule{0pt}{2.5ex}
					& \multirow{2}{*}{2p100} & \multirow{2}{*}{0.05} & \multirow{2}{*}{$2\times 10^{-4}$} & \multirow{2}{*}{$5\times 10^{-4}$} & \multirow{2}{*}{1.2, 1.95} & \multirow{2}{*}{10, 100} & \textcolor{cavitycolor}{29:20$^\ast$} \\
					&&&&&&& \textcolor{sigmacolor}{2:1}\\
					\rule{0pt}{2.5ex}
					& \multirow{2}{*}{2pJup} & \multirow{2}{*}{0.05} & \multirow{2}{*}{$2\times 10^{-4}$} & \multirow{2}{*}{$5\times 10^{-4}$} & \multirow{2}{*}{1.2, 1.95} & \multirow{2}{*}{10, 333}& \textcolor{cavitycolor}{2:1} \\
					&&&&&&& \textcolor{sigmacolor}{2:1}\\	
					\rule{0pt}{2.5ex}
					& \multirow{2}{*}{2p20LA} & \multirow{2}{*}{0.05} & \multirow{2}{*}{$2\times 10^{-4}$} & \multirow{2}{*}{$5\times 10^{-5}$} & \multirow{2}{*}{1.2, 1.95} & \multirow{2}{*}{10, 20}& \textcolor{cavitycolor}{unstable} \\
					&&&&&&& \textcolor{sigmacolor}{3:2}\\	
					\rule{0pt}{2.5ex}
					& \multirow{2}{*}{2p20LH} & \multirow{2}{*}{0.03} & \multirow{2}{*}{$2\times 10^{-4}$} & \multirow{2}{*}{$5\times 10^{-4}$} & \multirow{2}{*}{1.2, 1.95} & \multirow{2}{*}{10, 20}& \textcolor{cavitycolor}{4:3} \\
					&&&&&&& \textcolor{sigmacolor}{5:4}\\	
					\rule{0pt}{2.5ex}
					& \multirow{2}{*}{2p20HS} & \multirow{2}{*}{0.05} & \multirow{2}{*}{$4\times 10^{-4}$} & \multirow{2}{*}{$5\times 10^{-4}$} & \multirow{2}{*}{1.2, 1.95} & \multirow{2}{*}{10, 20}& \textcolor{cavitycolor}{5:4} \\
					&&&&&&& \textcolor{sigmacolor}{3:2}\\	
					\midrule
					\multirow{14}{*}{3} & \multirow{2}{*}{3p1020} & \multirow{2}{*}{0.05} & \multirow{2}{*}{$2\times 10^{-4}$} & \multirow{2}{*}{$5\times 10^{-4}$} & \multirow{2}{*}{1.2, 1.95, 3.2} & \multirow{2}{*}{10, 10, 20} & \textcolor{cavitycolor}{unstable} \\
					&&&&&&& \textcolor{sigmacolor}{3:2:1}\\	
					\rule{0pt}{2.5ex}
					& \multirow{2}{*}{3p2020} & \multirow{2}{*}{0.05} & \multirow{2}{*}{$2\times 10^{-4}$} & \multirow{2}{*}{$5\times 10^{-4}$} & \multirow{2}{*}{1.2, 1.95, 3.2} & \multirow{2}{*}{10, 20, 20} & \textcolor{cavitycolor}{unstable} \\
					&&&&&&& \textcolor{sigmacolor}{3:2:1}\\	
					\rule{0pt}{2.5ex}
					& \multirow{2}{*}{3p10100} & \multirow{2}{*}{0.05} & \multirow{2}{*}{$2\times 10^{-4}$} & \multirow{2}{*}{$5\times 10^{-4}$} & \multirow{2}{*}{1.2, 1.95, 3.2} & \multirow{2}{*}{10, 10, 100} & \textcolor{cavitycolor}{unstable} \\
					&&&&&&& \textcolor{sigmacolor}{3:2:1}\\	
					\rule{0pt}{2.5ex}
					& \multirow{2}{*}{3p20100} & \multirow{2}{*}{0.05} & \multirow{2}{*}{$2\times 10^{-4}$} & \multirow{2}{*}{$5\times 10^{-4}$} & \multirow{2}{*}{1.2, 1.95, 3.2} & \multirow{2}{*}{10, 20, 100}& \textcolor{cavitycolor}{unstable} \\
					&&&&&&& \textcolor{sigmacolor}{3:2:1}\\	
					\rule{0pt}{2.5ex}	
					& \multirow{2}{*}{3p2020LA} & \multirow{2}{*}{0.05} & \multirow{2}{*}{$2\times 10^{-4}$} & \multirow{2}{*}{$5\times 10^{-5}$} & \multirow{2}{*}{1.2, 1.95, 3.2} & \multirow{2}{*}{10, 20, 20} & \textcolor{cavitycolor}{unstable} \\
					&&&&&&& \textcolor{sigmacolor}{3:2:1}\\	
					\rule{0pt}{2.5ex}
					& \multirow{2}{*}{3p2020LH} & \multirow{2}{*}{0.03} & \multirow{2}{*}{$2\times 10^{-4}$} & \multirow{2}{*}{$5\times 10^{-4}$} & \multirow{2}{*}{1.2, 1.95, 3.2} & \multirow{2}{*}{10, 20, 20}& \textcolor{cavitycolor}{4:3:2} \\
					&&&&&&& \textcolor{sigmacolor}{5:4, 3:2}\\	
					\rule{0pt}{2.5ex}
					& \multirow{2}{*}{3p2020HS} & \multirow{2}{*}{0.05} & \multirow{2}{*}{$4\times 10^{-4}$} & \multirow{2}{*}{$5\times 10^{-4}$} & \multirow{2}{*}{1.2, 1.95, 3.2} & \multirow{2}{*}{10, 20, 20} & \textcolor{cavitycolor}{unstable} \\
					&&&&&&& \textcolor{sigmacolor}{3:2:1}\\	
					\midrule
					\multirow{4}{*}{>3} & \multirow{2}{*}{one-by-one} & \multirow{2}{*}{0.05} & \multirow{2}{*}{$2\times 10^{-4}$} & \multirow{2}{*}{$5\times 10^{-4}$} & \textcolor{cavitycolor}{1.2, 1.9, 2.43, 2.98} & \textcolor{cavitycolor}{10, 20, 20, 20} & \textcolor{cavitycolor}{unstable} \\
					&&&&&  \textcolor{sigmacolor}{1.2, 1.95 2.98, 4.49} & \textcolor{sigmacolor}{10, 20, 20, 20} & \textcolor{sigmacolor}{2:1, 2:1, --}\\
					\rule{0pt}{2.5ex}
					& \multirow{2}{*}{one-by-oneLH} & \multirow{2}{*}{0.03} & \multirow{2}{*}{$2\times 10^{-4}$} & \multirow{2}{*}{$5\times 10^{-4}$} & \textcolor{cavitycolor}{1.2, 1.76, 2.3, 2.8, 3.67, 4.48} & \textcolor{cavitycolor}{10, 20, 20, 20, 20, 20} & \textcolor{cavitycolor}{4:3, 3:2, 4:3, 5:4, --}\\
					&&&&&  \textcolor{sigmacolor}{1.2, 1.73, 2.19, 2.69} & \textcolor{sigmacolor}{10, 20, 20, 20} & \textcolor{sigmacolor}{unstable}\\
					\bottomrule
				\end{tabular}
				\caption{The list of our main simulations. The information about the \dz~ and \ib~ models is given by their corresponding colours. The data in black are the same in both trap models. \newline $^\ast$ The planets are very close to the 29:20 commensurability with $T_{\rm o}/T_{\rm i} \simeq 1.45$.}
				\label{tab:models}
			\end{table*}

		\section{Results}
		\label{sec:results}
		In this section we summarise the main results of our study. To analyse the effect of an inner edge on the final configuration of embedded planets, we ran an extensive suite of simulations using various disc parameters and planet masses. The set of models is summarised in Table.~\ref{tab:models}, and we refer to the quoted model identifiers in describing their outcome.
		The simulations were continued until all planet pairs settled into resonances or the system became unstable. As we found that the result of each simulation is specific, we present and explain them one by one in the following.
To make the whole text more readable, we display additional results on the individual models in Appendix~\ref{appendix:B}.
%
			\begin{figure}
				\centerline{\includegraphics[width=\columnwidth]{ 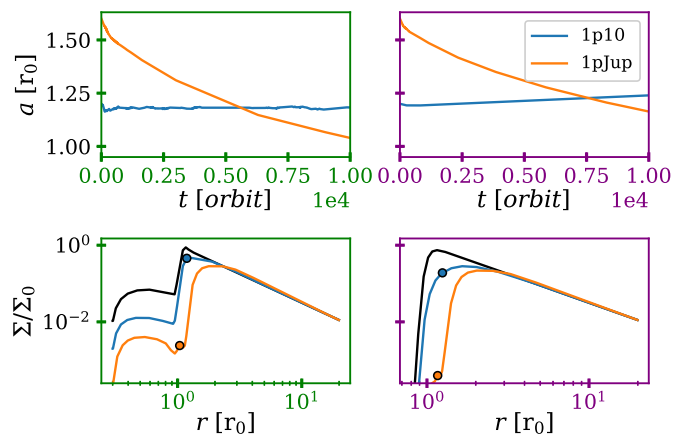}}
				\caption{ Evolution of a single planet in a disc with a \dz~(left) and at the \ib~(right). \textit{Top}: Migration path, $a(t)$. In each panel, every line belongs to a separate simulation with only one planet. \textit{Bottom}: Surface density profiles at $t=50$ (black), when the full potential of the planet is applied on the disc, and $t=10^{4}$~orbit. The colour code is the same as for the top row and the final locations of the planets are marked by circles. The low-mass planet is trapped the \dz~ model and slowly migrates outwards in the \ib~ model.
				In contrast, the massive planet (Jupiter mass) continues its migration to the inner disc because of the deformation of the bump by its gap.}
				\label{fig:singles}
			\end{figure}
%
		\subsection{Single planet}
		\label{sec:single}
			In the fist step, we checked the trapping of a single planet in our disc models.  The top panels of Fig.~\ref{fig:singles} show the migration of a $10\, \rm M_{\oplus}$ and a Jupiter-mass planet in a \dz~ (left) and a \ib~ (right) model (1p10 and 1p1Jup in Tab.~\ref{tab:models}). As pointed out, the colour of axes specifies the type of disc model. In the bottom rows, the surface density profiles are presented at $t=50$~orbits (in black), when the planet potential is fully applied on the disc, and $t=10^{4}$~orbits. 
			
			In the \dz~ model (with a steep density slope), the low-mass planet is trapped at the zero-torque location ($r\sim1.14$) and its semi-major axis does not evolve, as the torque map in Fig.~\ref{fig:discsetups} also suggests. The small oscillations in the semi-major axis is caused by some vortices outside of the edge. These vortices exert some torques on the planet as they pass by and produce small wiggles. In contrast, the massive planet continues its inward migration as it opens a planetary gap and deforms the density profile, which completely suppresses the positive co-rotation torque.
			
			In the \ib\,model, the low-mass planet moves slowly outwards as the trap regions widens and the zero-torque location moves slowly outwards. The massive planet in this model behaves similar to the \dz~ case and continues its inward migration by reshaping the surface density profile.
			
			These results imply that massive gap-opening planets might push and shepherd small planets that are trapped at the edge toward the star. In the next section, we add a second planet with various masses to these models and also examine the effect of lower viscosity and lower aspect ratio.
%
			\begin{figure}
				\centerline{\includegraphics[width=\columnwidth]{ 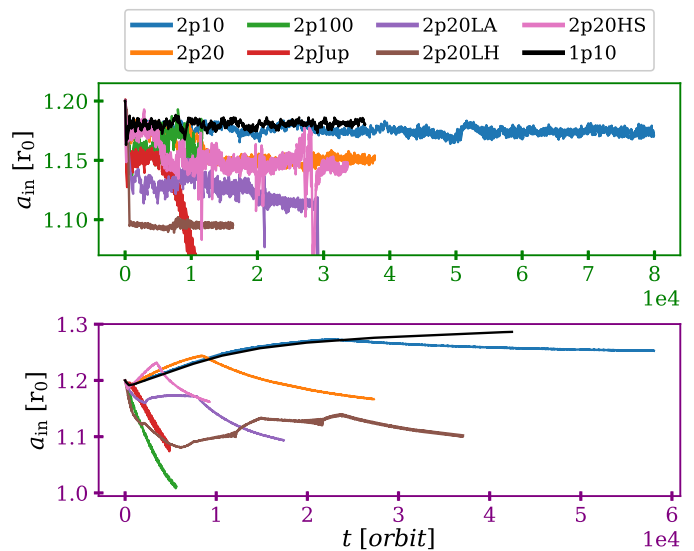}}
				\caption{Semi-major axis evolution of the inner $10\, \rm M_{\oplus}$ planet for all models with two planets. Top for DZ models and bottom for the IB models. The single-planet model (1p10) is added for comparison.}
				\label{fig:2pallain}
			\end{figure}
%
		\subsection{Planet pair}
		\label{sec:pair}
			In this section, we add a second planet just off the 2:1 commensurability and monitor the evolution of the system. The mass of the inner planet in all cases was $10\, \rm M_{\oplus}$, but the outer planet had different masses. We denote the inner planet by the index $i$ and the outer planet by $o$.
			Taking the case with $M_\mathrm{o}=20\, \rm M_{\oplus}$ (model 2p20) as the reference model, we additionally examined the effect of lower viscosity, lower aspect ratio, and higher surface density for this model. 
			
			Figure.~\ref{fig:2pallain} shows the migration of the inner planet for all models with two planets. The top panel contains the results for the \dz~ model (green frame) and the bottom panel (purple frame) shows those of the \ib~ model. The figure clearly shows that the second planet undoubtedly has an effect on the migration of the inner planet, especially if it is more massive than the inner planet. Moreover, when we compare the top and bottom panels, it is apparently easier to push the inner planet to the inner disc in the \ib~ models than the \dz~ models. In the \dz~ models, the inner planet persists at the trapping location, while in the \ib~ models, the outer planet in most cases is able to reverse the migration of the inner planet and pushes it inward after they settle into resonance. In the following, we describe the results of the models individually in more detail.

\begin{figure}
				\centerline{\includegraphics[width=\columnwidth]{ 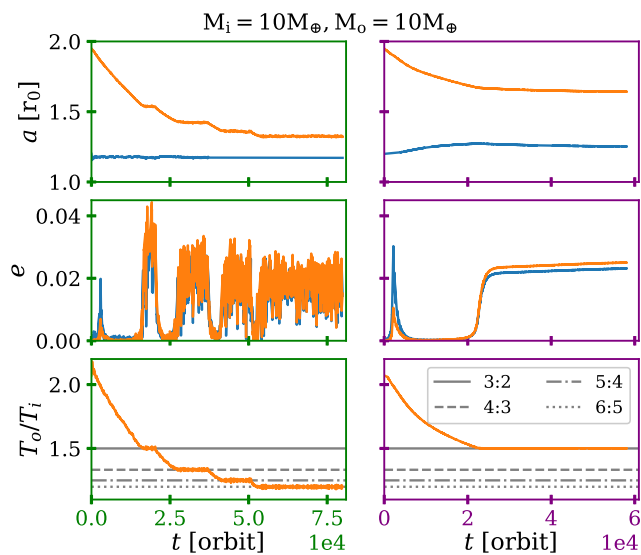}}
				\caption{Evolution of the semi-major axis, eccentricity, and orbital period ratio for 2p10 models. The inner planet is shown by the blue line and the outer planet by the orange line. In the bottom panels resonances are marked by horizontal lines.}
				\label{fig:21010}
			\end{figure}

\begin{figure}
				\centerline{\includegraphics[width=\columnwidth]{ 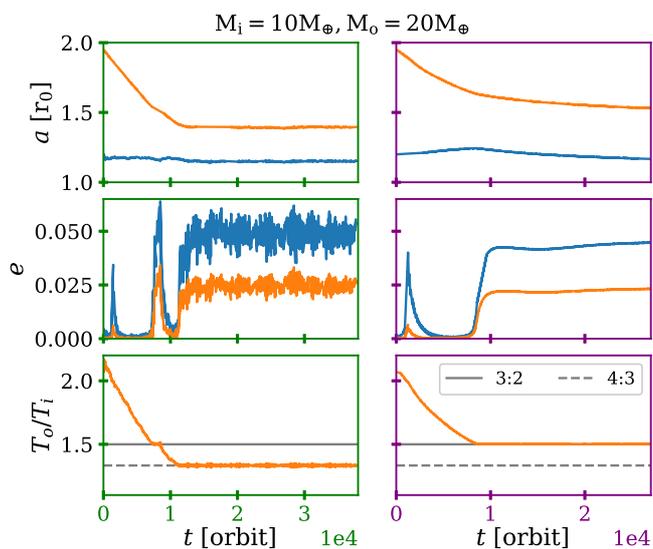}}
				\caption{Similar to Fig.~\ref{fig:21010}, but for the 2p20 model.}
				\label{fig:21020}
			\end{figure}

			\textit{2p10}: In these models, the masses of both planets are $10\, \rm M_{\oplus}$. As Fig.~\ref{fig:21010} shows, in the \dz~ model, the planets cross several resonances and finally settle in the 5:4 resonance. The blue line in the top panel of Fig.\,\ref{fig:2pallain} indicates that the inner planet remains almost at its initial trapping location. In this case, the push by the outer planet is not strong enough to overcome the positive torque on the inner planet. In contrast, in the \ib~ model, after the planets reach the 3:2 resonance, the outer planet pushes the inner planet inwards and reverses its slow outward migration.
%
%
			\textit{2p20}: Figure~\ref{fig:21020} shows the semi-major axis, eccentricity, and orbital period ratio for 2p20 models. In the \dz~ simulation, the outer planet migrates inwards and reaches the 4:3 resonance with the trapped inner planet. Because the outer planet is twice as massive as the 2p10 model, the inner planet is pushed slightly into the transition zone (top panel of Fig.~\ref{fig:2pallain}). After forming the final resonance, the migration of the planets stops. In contrast, in the \ib~ simulation, after the outer planet catches the outwards-migrating inner planet in the 3:2 resonance, they continue their migration inwards, but at a slower rate than the rate of outer planet before the resonance. 
			
\begin{figure}
				\centerline{\includegraphics[width=\columnwidth]{ 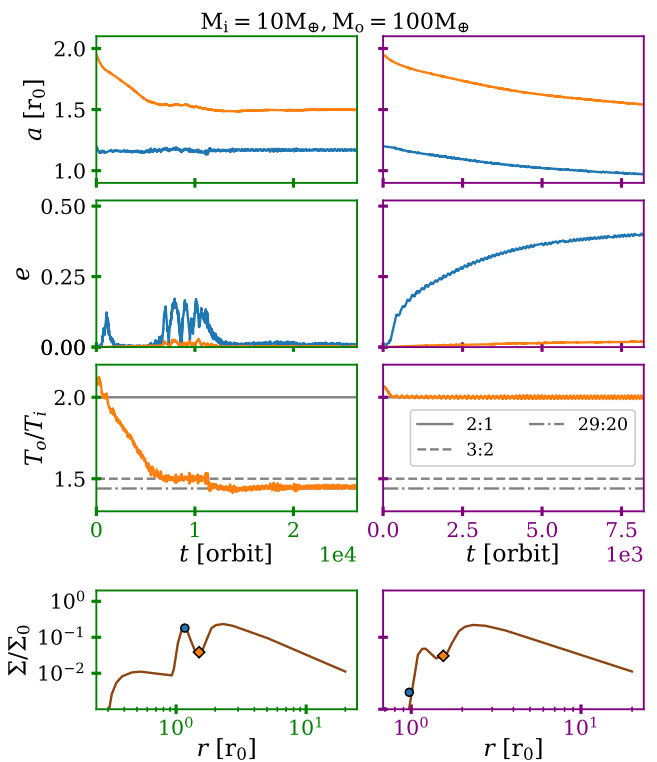}}
				\caption{Top rows are the same as Fig.~\ref{fig:21010}, but for the 2p100 model. The last row shows the final surface density. The locations of the planets are indicated by markers, whose colours correspond to the top panels.}
				\label{fig:2p100}
			\end{figure}

			\textit{2p100}: Fig.~\ref{fig:2p100} shows the results for an outer planet of 100\,M$_\oplus$. In addition to the previous figures, the bottom panel displays the surface density of the disc at the end of simulations. Locations of planets are indicated by markers. The outer planet in these models opens a partial gap, not as strong as the gap of the Jupiter-mass planet in 1pJup to destroy the bump, but can still affect it. The left panels in this figure that contain the results for the \dz~ model, show that the outer planet cages the inner planet between the edges of the dead-zone and its gap. After temporarily settling into the 3:2 MMR, they rearrange and settle near a 29:20 commensurability. We checked the resonant arguments but found no definite pattern. In the \ib~ model (right panels), the planets reach the 2:1 resonance and maintain this configuration until the end of the simulation. Both planets continue their inward migration and the inner planet, which becomes very eccentric with the eccentricity of $\sim 0.4$, is pushed to the inner part of the disc.

			\textit{2pJup}: In these models the outer planet has one Jupiter mass and opens a gap similar to the 1pJup model. For both edge models, the inner planet is trapped in the 2:1 resonance with the outer one (see Fig.~\ref{fig:2101Mj} in Appendix~\ref{appendix:B}). The massive planet pushes the inner planet into the inner part of the disc regardless of which inner disc model is used. The low-mass planet in both models becomes very eccentric. We stopped the simulations because for such high eccentricities $\sim 0.3$, the inner planet passes the area that is close to the inner boundary of our computational domain. The overall behaviours of the planets in both of these models are very similar and it seems that the type of the trap has only little effect on the evolution. 
			Therefore we did not examine more models with a Jupiter-mass planet.
			
\begin{figure}
				\centerline{\includegraphics[width=\columnwidth]{ 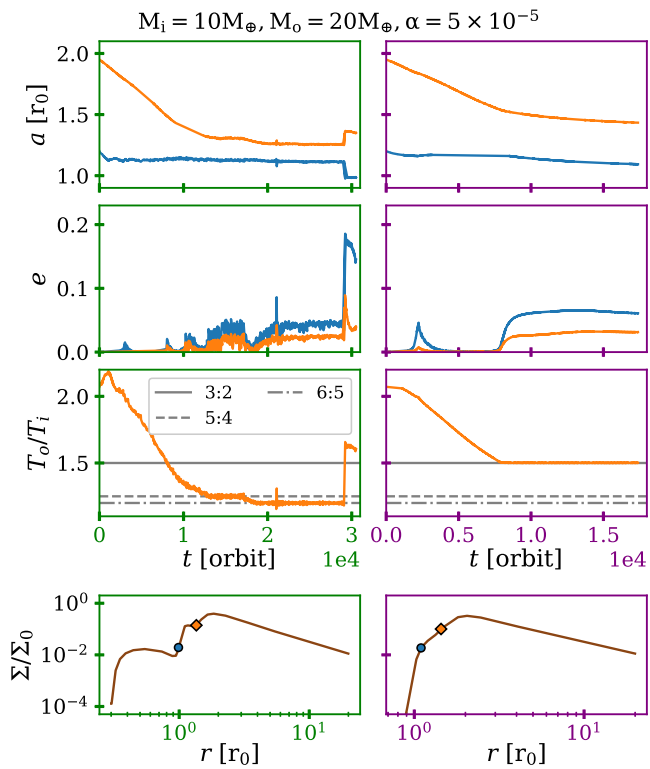}}
				\caption{Similar to Fig.~\ref{fig:2p100}, but for 2p20LA models that have a reduced viscosity.}
				\label{fig:2p20LA}
			\end{figure}

			\textit{2p20LA}: These models have identical setups as the reference 2p20 models except that $\alpha$ is lowered to $5\times 10^{-5}$. The lower viscosity can facilitate the partial gap opening and consequently reduce both of the Lindblad and co-rotation torques. In addition, it can  change the saturation of the co-rotation torque \citep{2001ApJ...558..453M,2012ARA&A..50..211K}. In the bottom panels of Fig.~\ref{fig:2p20LA} the surface density profiles at the end of the simulations are displayed. We can distinguish a shallow gap around the outer planet with the mass of $20\, \rm M_{\oplus}$. Comparing the two upper rows of Fig.\ref{fig:2p20LA} with those of Fig.\ref{fig:21020} shows that the planets in the \dz~ model managed to approach and settle in the 6:5 resonance. However, they finally closely interact, and the inner planet is sent to an orbit outside of the bump, and the order of the planet is reversed. A comparison with the top panel in Fig.~\ref{fig:2pallain} shows that the inner planet here is initially trapped farther inside than the planet in the 2p20 and 1p10 models with higher viscosity. This is consistent with the idea that the lower viscosity reduces the positive co-rotation torque through the saturation effects.
			
			In the \ib~model, similar to its 2p20 counterpart, the planets pass the 2:1 resonance (around $\sim 2000$~orbits when their eccentricities peak), reach the 3:2 resonance, and after this, the inner planet is pushed to the inner disc. A closer look at the semi-major axis of the inner planet (lower panel of Fig.~\ref{fig:2pallain}) shows that in contrast to the 2p20 model, the inner planet initially migrates inwards, which indicates that the co-rotation torque is more saturated. After the outer planet passes the 2:1 resonance, the inner planet changes its migration direction, moves outwards, and halts until the outer planets catch it in the 3:2 resonance. Thereafter, its migration is governed by the outer planet.
			
			\textit{2p20LH}: These models have a lower aspect ratio than 2p20, here $h=0.03$. A lower aspect ratio can change the torque in two ways. Firstly, the torque from the disc on the planet is proportional to $h^{-2}$ \citep[][and references therein]{2014prpl.conf..667B}, and therefore the torque is stronger for smaller $h$. Secondly, the planets are liable to open a partial gap, and the smaller surface density around the planetary orbit decreases the torque. The brown lines in Fig.~\ref{fig:2pallain} show that the semi-major axes of the inner planets for the 2p20LH models are smaller than those of the reference model 2p20. 
%
			
\begin{figure}
				\centerline{\includegraphics[width=\columnwidth]{ 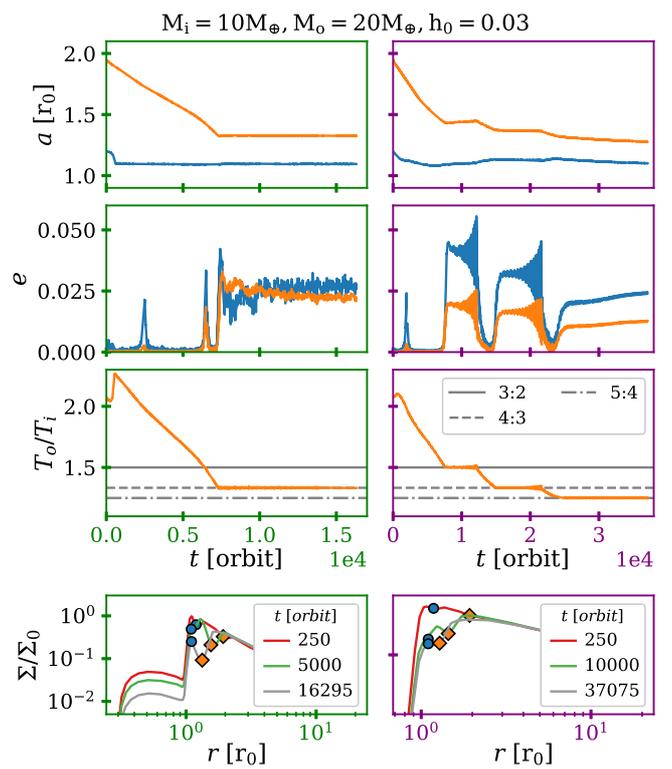}}
				\caption{Similar to Fig.~\ref{fig:2p20LA}, but for 2p20LH except that the panels of the last row show three different times to demonstrate the gap opening of the planets. The markers display the location of each planet, and their colours correspond those in the upper panels.}
				\label{fig:21020LH}
			\end{figure}

			The left panels of Fig.~\ref{fig:21020LH} show that the planets in the \dz~ model finally settle into the 4:3 resonance and remain there.
			In the \ib~ model, whose results are displayed in the right panels of Fig.~\ref{fig:21020LH}, the inner planet migrates inwards until outer planet captures it in the 3:2 resonance at around $t\approx 7500$~orbit. As long as they are in this configuration, they migrate very slowly outwards until they leave the resonance. A similar behaviour occurs later, when they reach  the 4:3 resonance. Finally, they leave the 4:3 resonance and eventually stay in the 5:4 resonance while they continue their inward migration. Their semi-major axis and particularly their eccentricity evolutions show an over-stable behaviour in the 3:2 and 4:3 resonances. We discuss about over-stable models in Sec.~\ref{sec:overstability}.
		
			\textit{2p20HS}: The surface density in these models is twice that of the 2p20 models. Hence, the planets feel stronger torques than in 2p20. The migrations and eccentricities are almost identical to the 2p20 model, but the evolution is faster. The lower panel of Fig.~\ref{fig:2pallain} shows that in the \ib~ model the behaviour of the inner planet is very similar in the 2p20 and 2p20HS models, but the resonance capture occurs earlier in 2p20HS. However, the results are slightly different for the \dz~ model. The planets are initially reach and stay in 5:4 resonance for some time, then are pushed into and remain in the deeper 6:4 resonance for about \num{10000} orbits. After this, they return to the 5:4 resonance because of the planet-planet gravitational interaction (see Fig.~\ref{fig:21020HS} in the Appendix~\ref{appendix:B}). Although the resonance configuration changes, the semi-major axis of the inner planet does not vary and remains the same as in the 2p20 model after the first resonance (upper panel of Fig.~\ref{fig:2pallain}).

		\subsection{Three planets}
		\label{sec:threeplanets}
\begin{figure}
				\centerline{\includegraphics[width=\columnwidth]{ 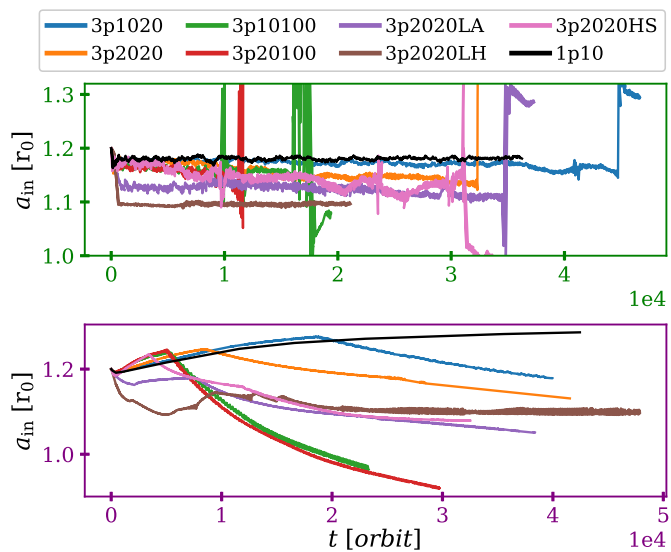}}
				\caption{Semi-major axis evolution of the innermost planet. Similar to Fig.~\ref{fig:2pallain}, but for models with three planets. The inner planet always has $10\, \rm M_{\oplus}$, and the labels give the masses of the other two in units of $\rm M_{\oplus}$, see Tab.~\ref{tab:models}.}
				\label{fig:3pallain}
			\end{figure}

			In this section, we present the results of the simulations with three planets. The mass of the outer planet was  either $20$ or $100\, \rm M_{\oplus}$, the middle masses were $10$ or $20\, \rm M_{\oplus}$, and the inner planet was always $10\, \rm M_{\oplus}$, see Tab.~\ref{tab:models} for details. The outer planet in these models was added just outside of 2:1 commensurability with the middle planet. Figure~\ref{fig:3pallain} shows the migration of the inner planet for all models with three planets. In the top panel, which shows the results for the \dz~ models, the inner planet in many simulations undergoes an instability and is expelled. However, none of the inner planets is ejected in the \ib~ model (the bottom panel of Fig.~\ref{fig:3pallain}). In the following, we describe the results of each model individually.

			\begin{figure}
				\centerline{\includegraphics[width=\columnwidth]{ 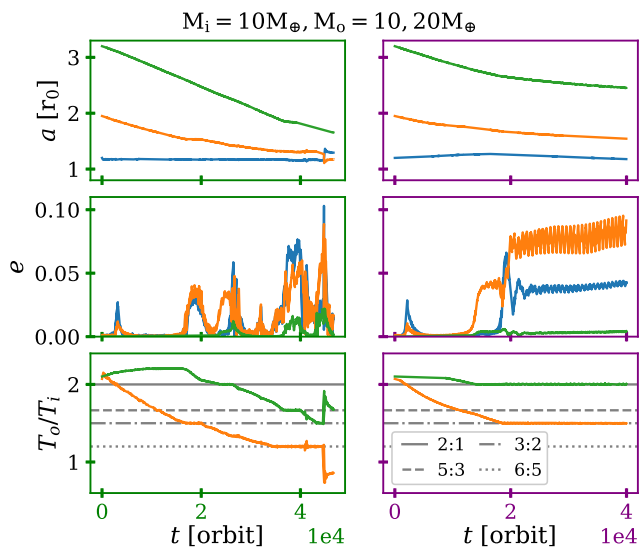}}
				\caption{Semi-major axis and eccentricity for 3p1020 models (top and middle panels). The bottom panels show the orbital period ratio for each consecutive pair of planets, where the colour corresponds to the outer planet. For example, the green curve demonstrates the orbital period ratio of the third planet to the second planet from the inside out.}
				\label{fig:3101020}
			\end{figure}
			
			\textit{3p1020}: These simulations have the same setup as 2p10, but a third planet with a mass of $20\,\rm M_{\oplus}$ was added. With the aid of the black lines in Fig.~\ref{fig:2pallain} and \ref{fig:3pallain}, which show the semi-major axis of a $10 \rm M_{\oplus}$ planet in the single-planet models, we can compare the effect of adding the third planet to the evolution of the inner planet. Comparison of the blue lines in the upper panels of the mentioned figures indicates that in the \dz~ model, the presence of the third planet hardly affects the migration of the inner planet. When the third planet catches the other two in a 6:5:3 resonant chain (with period ratios 6:5 and 5:3 from the inside out) at $\sim 35000$~orbits (see the left panels of Fig.~\ref{fig:3101020}), the inner planet is pushed slightly inwards. The outer planet later leaves the chain and continues its inward migration until the two inner planets closely interact and switch their orbits. In the \dz~ model, the outer planet therefore pushes the system into a tighter configuration, which is more prone to instability. 
			
			In the \ib~ model, the outer planet forms the 2:1 resonance with the middle planet, but has no effect on the migration of the inner planet (right panels of Fig.~\ref{fig:3101020}). As the bottom panel of Fig.\ref{fig:3pallain} shows, the semi-major axis of the inner planet follows that of the single planet until it retains the 3:2 resonance with the middle planet. Afterwards, it is pushed into the inner disc with a faster migration rate than its 2p10 counterpart.
			
			\textit{3p2020}: The trapping positions of the inner planets in these models are very similar to their two-planet counterparts, and the outer planet does not play a notable role. In the \dz~model, although the outer planet does not push the innermost planet, it can destabilise the system after it enters the 2:1 resonance with the middle planet~(Fig.~\ref{fig:3102020}). After the two outer planets spend some time in this resonance, the inner and middle planets switch orbits. In the \ib~ model the middle planet enters the 3:2 resonance  with the inner planet at around 8000 orbits, and it causes the eccentricity of the inner planet to increase, exactly as in the 2p20 model, see Fig.\,\ref{fig:21020}. The outer planet later catches up, and a 3:2:1 resonant chain forms with period ratios 3:2 and 2:1 (from the inside out). Upon the capture, the eccentricities of the two inner planets increase further. Otherwise, no change in the migration rate of the inner planet occurs.
		
			\begin{figure}
				\centerline{\includegraphics[width=\columnwidth]{ 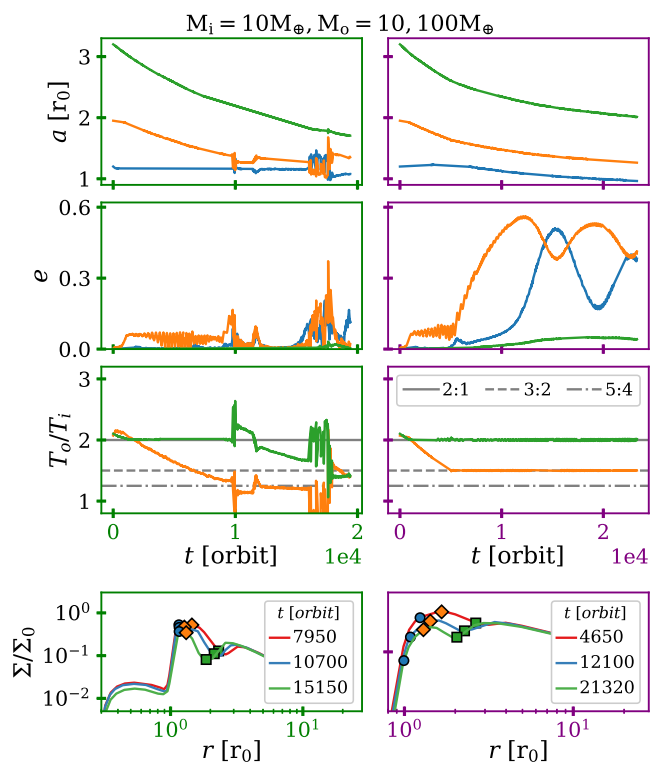}}
				\caption{Similar to Fig.~\ref{fig:3101020}, but for the 3p10100 model.}
				\label{fig:31010100}
			\end{figure}
			
			\textit{3p10100}: These models are complementary to the 2p10 and 3p1020 models in understanding the effect of the third planet. Comparison of the \dz~simulations in the three mentioned models, shown in the left panels of Fig.~\ref{fig:21010}, \ref{fig:3101020}, and \ref{fig:31010100}, indicates that the two inner planets form tight configurations in all three models, and the role of the outer planet is mostly to destabilise the system. The more massive the outer planet, the earlier the system becomes unstable. Differently, in the \ib~ simulations, the third planet has a notable effect. When the third planet is more massive, the reversal of the inner planet migration occurs earlier, and the inward migration is faster after the reversal. In all of these \ib~ models, the two inner planets are in 3:2 resonance, and when a third planet exists, the planets form a 3:2:1 resonant chain, which enhances the eccentricity of the lower mass planets. 
			
			The last row of Fig.~\ref{fig:31010100} shows the azimuthally averaged surface density of the 3p10100 models at specific times. In both cases, the outer planet opens a gap. In the \dz~ model, the two inner planets are sandwiched between the dead-zone edge and the partial gap of the outer planet. In contrast to the \dz~ model that becomes unstable, the planets in the \ib~ model migrate inwards while keeping their resonant chain. 
			
			\textit{3p20100}: The evolution of these models are very similar to 3p2020, but are faster because the migration of the more massive outer planet is faster. Similar to 3p2020, the \dz~model becomes unstable after about $\num{10000}$~orbits.
			The planets in the \ib~ model form a 3:2:1 chain. The only difference is that the eccentricities of the planets in this configuration are higher than those of 3p2020, because the mass of the outer planet is higher. The migration of the inner planet for this model and 3p10100 in the lower panel of Fig.~\ref{fig:3pallain} indicates that the behaviour of the inner planet after reversal is determined completely by the third planet, which is more massive than the second one. (The eccentricity and orbital period ratio for these models are shown in Fig~\ref{fig:31020100}.)
			
			\textit{3p2020LA}: These models with a smaller $\alpha$ behave very similar to their high-viscosity counterparts~(as shown in Fig.~\ref{fig:3102020LA}). The \dz~ model becomes unstable when the outer planet approaches the inner two, which are in 6:5 resonance. The planets in the \ib~ model, similar to the previous \ib~ models with three planets, continue their inward migration as they stay in the 3:2:1 resonant chain.
		
			\begin{figure}
				\centerline{\includegraphics[width=\columnwidth]{ 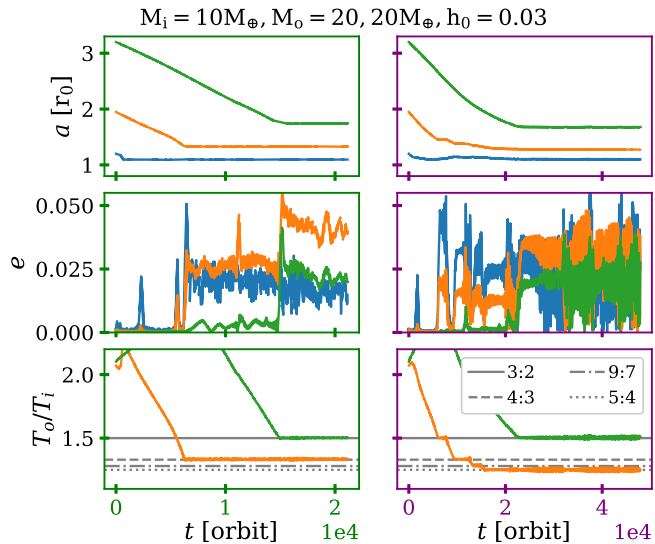}}
				\caption{Evolution of eccentricity and orbital period ratio for 3p2020LH, which has a smaller disc thickness.}
				\label{fig:3102020LH}
			\end{figure}
			
			\textit{3p2020LH}: These models, with a lower aspect ratio, show different behaviours than the other three-planet models. The \dz~ model here is the only one of the other \dz~models with three planets that does not become unstable after being ina 4:3:2 chain for more than \num{5000}~orbits. Because the inner planet barely moved from its trapping location (upper panel of Fig.\ref{fig:3pallain}), the migration of the whole system halted as they formed the resonant chain. This is shown in the left panels of Fig.~\ref{fig:3102020LH}. In the \ib~ model, as the lower panel of Fig.\ref{fig:3pallain} shows, the inner planet passes various phases of inward and outward migration depending on in which resonance it is with the outer planet. Even after the final configuration forms, it does not migrate inwards. The right panels of Fig.\ref{fig:3102020LH} demonstrate the evolution of this system. The two inner planets enter into various resonances, namely 3:2, 4:3, and 9:7, but these resonances are broken by the over-stability. They finally dwell in 5:4 resonance. The outer planet, however, remains in the 3:2 resonance with the middle planet after they enter into it.
		
			\textit{3p2020HS}: The outcome of these models with higher surface density and 3p2020 models are alike, except that the ones here evolve faster. Therefore we omit showing the full results. Fig.~\ref{fig:3pallain} shows that the inner planets in these models behave very similarly to their low surface density counterparts, but their inward movements have started earlier. As before, the \ib~ model ends up in the 3:2:1 chain and the \dz~ model becomes finally unstable. 
	
		\subsection{More than three planets}
		\label{sec:moreplanets}
			In this section, we examine the effect of additional planets in the systems and their effect on the innermost planet. We took the 2p20 as the reference model, add more planets with the mass of $20\, \rm M_{\oplus}$, and observed the behaviour of the innermost planet. We started with the inner $10\, \rm M_{\oplus}$ planet and allowed it to find its trapping location. Then the second planet with a mass of $20M_{\oplus}$ was placed just out of 2:1 commensurability. We continued the simulation until the planets were trapped in a resonance. The next planets were added one by one in a similar way when the inner pair formed a resonance.  We continued until the system became unstable or we found that adding more planets had no effect on the location of the inner planet. In the following, the results of these simulations for discs with $h=0.05$ and $h=0.03$ for the two inner edge models are presented.
			
			\begin{figure}
				\centerline{\includegraphics[width=\columnwidth]{ 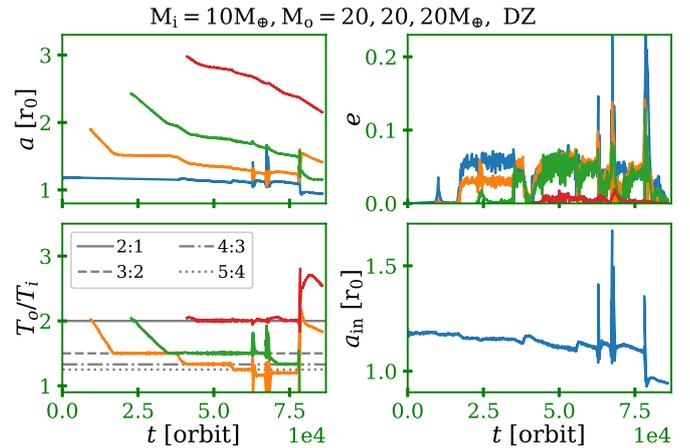}}
				\caption{Results of the one-by-one simulation for the \dz~ model. The top row displays the semi-major axes (left) and eccentricities (right) of all planets. The bottom left panel contains the orbital period ratio for each neighbouring pair. In the bottom right panel, the migration of the innermost planet is shown for more clarity.}
				\label{fig:oneonecav}
			\end{figure}
			
			\begin{figure}
				\centerline{\includegraphics[width=\columnwidth]{ 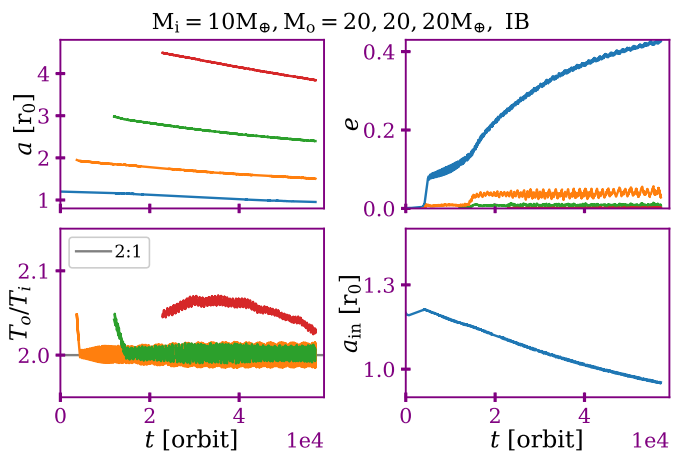}}
				\caption{Similar to Fig.~\ref{fig:oneonecav}, but for the one-by-one \ib~ model.}
				\label{fig:oneonesig}
			\end{figure}
			
			\textit{One-by-one}: The evolution of the system for the \dz~ model is shown in Fig.~\ref{fig:oneonecav}. In the top left panel, the semi-major axes of the planets are plotted. This panel shows in addition to the migration of the planets, where and when the planets were added into the disc. As the top right panel shows, the eccentricities are about \num{0.07} unless when the planets interact closely. With the help of the bottom panels, we can determine how the location of the inner planet is affected by the outer planets. The second planet migrates inwards until it forms 3:2 resonance with the first planet. The bottom right panel reveals that the inner planet is hardly pushed inwards at this time ($t \sim \num{14000}$~orbit). Shortly after the third planet forms a 3:2 resonance with the second planet, the two inner planets go to the tighter 4:3 resonance. The forth planet remains in 2:1 resonance with the third planet, and the whole system forms a 4:3:2:1 chain. As long as the system abides in this configuration, the inner planet moves slowly inwards. Then the two inner planets enter the tighter 5:4 resonance and the system becomes unstable thenceforth. This instability sends the inner planet out of the transition zone into the inner disc.
			
			The \ib~ model, unlike the \dz~ simulation, evolves very smoothly. The two inner pairs reach 2:1 resonance and remain in it. Thereafter, the migration of first planet reverses and becomes inwards. Later, its migration rate is not affected by the presence of the other planets. The third planet, however, ignites the eccentricities of the first and second planet after it forms 2:1 resonance with the second planet. The last planet seems to have no major effect on the system. We did not continue this simulation because of the high eccentricity of the inner planet.
			
			\begin{figure}
				\centerline{\includegraphics[width=\columnwidth]{ 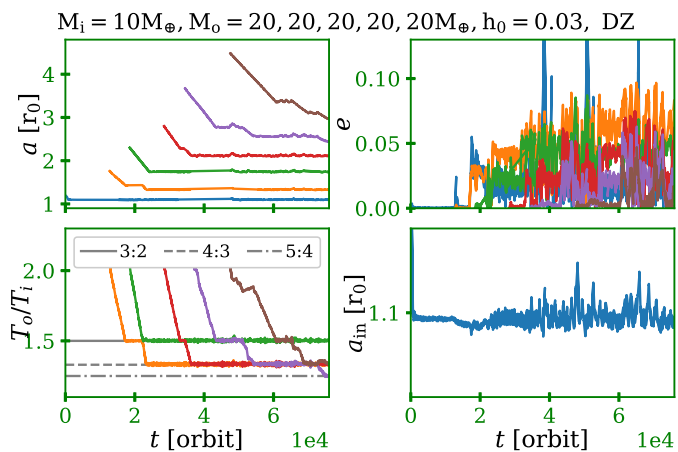}}
				\caption{Similar to Fig.~\ref{fig:oneonecav} but for the one-by-oneLH model which has a lower aspect ratio.}
				\label{fig:oneoneLHcav}
			\end{figure}
			
			\begin{figure}
				\centerline{\includegraphics[width=\columnwidth]{ 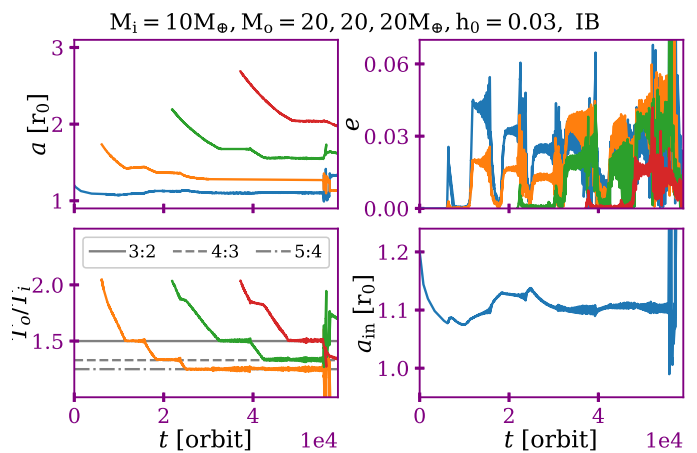}}
				\caption{Same as Fig.~\ref{fig:oneonesig} but for \ib\, edge model.}
				\label{fig:oneoneLHsig}
			\end{figure}
			
			\textit{One-by-oneLH}: In these simulations the lower disc aspect ratio changes the migration rate compared to the two previous cases. Unlike the one-by-one models, the \dz~ model here remains stable even with six planets, while the \ib~ model becomes unstable after the fourth planet. As Fig.~\ref{fig:oneoneLHcav} shows, we added five $20\, \rm M_{\oplus}$ planets to the system in the \dz~ model. All of them remained in either 3:2 or 4:3 resonances but the inner planet was not pushed inwards. The only effect of the outer planets on the first planet is to produce some jitters in its semi-major axis by enhancing its eccentricity. 
			
			The \ib~ model, on the other hand, as Fig.~\ref{fig:oneoneLHsig} demonstrates, becomes unstable after the fourth planet comes into resonance with the third planet and forms a 5:4:3:2 resonant chain. From the top right panel, we can infer that the two inner pairs undergo over-stability. The inner planet shows absolutely no monotonic behaviour. Depending on which resonance it forms with the other planets, it can move inwards or outwards.
		
	\section{Transition width and vortices}
	\label{sec:transwidth}
		For our choice of the viscosity transition in the \dz~ models, some vortices were created and persisted during the simulations. They form because of the narrow transition width and a small discontinuity in $\alpha$ at the edges of the transition zone. In this section, we examine different widths and profiles in order to inspect whether the vortices and transition width can affect the migration and final settling of the planets. For models 2p20 and 3p2020, we ran additional simulations using different viscosity profiles, either by changing $\lambda$ in Eq.~(\ref{eq:visc}) or by smoothing the transition.

		\begin{figure}
			\centerline{\includegraphics[width=\columnwidth]{ 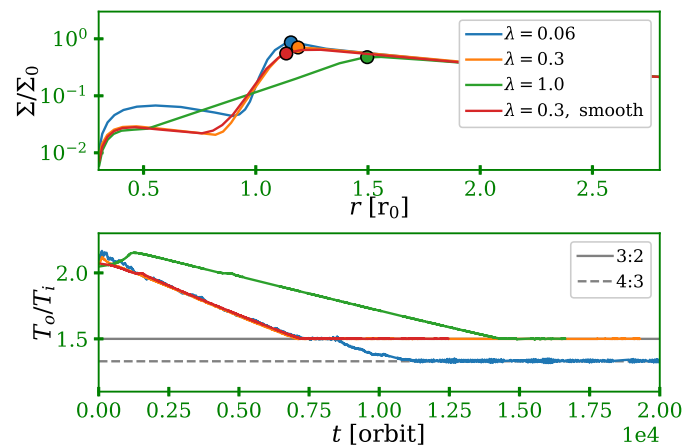}}
			\caption{Results for the 2p20 model using different transition widths for the \dz~case.
			\textit{Top}: Azimuthally averaged surface density of the inner part of the discs with different transition widths, $\lambda$, see Eq.~(\ref{eq:visc}). The profile labelled with $\lambda=0.3\ smooth$ has a transition width of $\lambda=0.3$ but the viscosity transition function is smoother at the edges than in our standard model. The final location of the inner planet is marked by a circle for each model. \textit{Bottom}: Orbital period ratio for the models in the top panel plotted with the corresponding colours.}
			\label{fig:cavwidthtratio}
		\end{figure}

		The top panel of Fig.~\ref{fig:cavwidthtratio} shows four azimuthally averaged surface density profiles for models with a viscosity profile given by relation~(\ref{eq:visc}) using $\lambda=0.06, 0.3, 1.0$, and one that has the same transition width as $\lambda=0.3$, but using a smoother function at the edges of the transition zone. The bottom panel of Fig.~\ref{fig:cavwidthtratio} presents the orbital period ratio for these simulations. In all of them, the planets retain 3:2 resonance, except for the planet with the narrowest transition width $\lambda=0.06$ (our reference model), for which the outer planet was able to migrate closer to the inner planet. 

\begin{figure}
			\centerline{\includegraphics[width=\columnwidth]{ 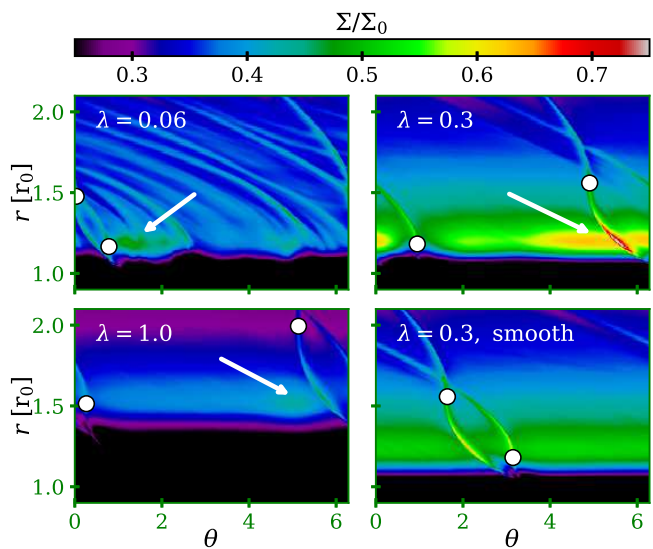}}
			\caption{Two-dimensional surface density for the discs in Fig.~\ref{fig:cavwidthtratio}. Planets are marked with white circles, and the arrows show the centre of the (largest) vortex in the models with a vortex or several vortices.}
			\label{fig:cavitywidth2dsig}
		\end{figure}

		The corresponding 2D surface densities are displayed in Fig.~\ref{fig:cavitywidth2dsig} at the time just before the planets reach the resonance. The three simulations with our standard profile, show at least one vortex, while the simulation with the smoother transition shows none. In all of these simulations, regardless of the existence of a vortex or several vortices, the inner planet is trapped at the edge of the transition zone. 
		The model with the steepest transition, $\lambda=0.06$, shows many small vortices at the transition edge. Although they can hardly be distinguished in Fig.~\ref{fig:cavitywidth2dsig}, their spirals are clearly visible \citep{2010ApJ...725..146P}. Both the vortices and their spirals can interact with the planets and produce small perturbations in their orbits. However, because these interactions are random, their torques or powers on the planets do not affect the overall migration. For the models with the wider transitions $\lambda=0.3$ and $\lambda=1$, only a single vortex is established and the inner planet is trapped azimuthally between the two vortex endpoints \citep{2014A&A...572A..61A}. In all of these models, after the planets maintain the resonance and their eccentricities are excited, the vortex or vortices fade.
		
		In the model with the smooth transition (labelled $\lambda=0.3\ \rm smooth$), as shown in Figs.~\ref{fig:cavwidthtratio} and \ref{fig:cavitywidth2dsig}, the surface density profile is almost identical to the model with $\lambda=0.3$, but no vortex is created during the evolution. In both of these models (smooth and not smooth) the planets are trapped in 3:2 resonance. However, the inner planet in the model with smooth transition is stopped slightly farther inwards than the one with the standard profile. These additional tests show that the vortices near the edge do not drastically alter the results, but they create some noise in the simulations.
		
		The planets in the 3p2020 simulations with $\lambda = 0.3$ and $1.0$ are trapped in 3:2 resonance and do not become unstable during the simulation time. In spite of the model with $\lambda=0.06$, where the system reaches 5:4 resonance and becomes unstable, the inner planet in the simulation with $\lambda=0.3$ is trapped at the transition edge, while in the widest model ($\lambda=1$), the outer planet pair managed to push the inner planet out of the trapping point (Fig.~\ref{fig:cavwidth3ptratio}). Therefore it appears that similar to the \ib~models, the inner planet can be pushed into the inner disc if the transition zones in the \dz~models are extremely wide.

	\section{Power and torque analysis of the 2p20 models}
	\label{sec:torpow}
		The results of two-planet models show that after the resonance formation, the inner planet in the \dz\ simulations is slightly pushed inwards, where the torque is expected to be positive. After that the migration of the whole system halts. In the \ib\ models, in spite of the initial outward migration of the inner planet, the resonant system pushes the inner planet back to the inner disc and the whole system migrates inwards with a different migration rate than those of the individual planets. In this section we analyse the torque and power for 2p20 model (displayed in Fig.~\ref{fig:21020}), and show that the halting position of the whole system is at the location where the total power vanishes.
		
		\begin{figure}
			\centerline{\includegraphics[width=\columnwidth]{ 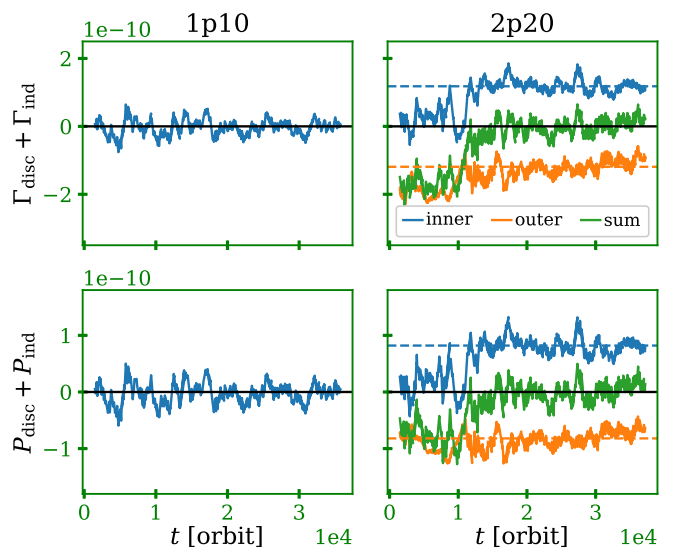}}
			\caption{ Total torque (top row) and total power (bottom row). These quantities for a single planet of the 1p10 \dz~ model are shown in the left column and for the 2p20 model are shown in the right column. The dark green line represents the total torque or power on the system.
			The horizontal black lines in each panel marks zero. The contribution of the indirect term is included in the plotted quantities. Torques and powers are averaged over 1000 orbits to decrease the oscillations created by the vortices and eccentricities of the planets. The dashed lines in the right panels correspond to the values of the eccentricity and semi-major axis damping timescales used in the supporting N-body simulations, as displayed in Fig.~\ref{fig:nbodycav}.}
			\label{fig:torqpowcav}
		\end{figure}
		
		The energy $E$, and angular momentum $L$, of a planet is given in terms of its semi-major axis and eccentricity as
		\begin{equation}
             \label{eq:energy-angmom}
		     E= - \frac{GM_{\star}M_{\rm p}}{2a}  \quad  \mbox{and} \quad  L=M_{\rm p} \sqrt{ G M_{\star} a ( 1 - e^2 )} \,.
		\end{equation}
		As the energy depends only on the semi-major axis of the planet, its migration rate is determined by the disc power, $P_{\rm disc}$, acting on it. The power gives the total energy change of a planet
		\begin{equation}
			\label{eq:dotE}
             \dot{E}  = P_{\rm disc} \,.
		\end{equation}
		On the other hand, the disc torque, $\Gamma_{\rm disc}$ determines the rate of angular momentum change of a planet
		\begin{equation}
			\label{eq:dotL}
             \dot{L}  = \Gamma_{\rm disc} \,,
		\end{equation}
		which is a combination of eccentricity and semi-major axis change \citep[see e.g.][]{2007A&A...473..329C}.
		For a single planet on a circular orbit, torque and power are equivalent. In this case, it is expected that the migration of planet halts where the torque vanishes. However, for a planet on an eccentric orbit, the halting location will be given by vanishing power. In order to analyse the halting process of a planet pair, we monitored the disc torque and power continuously during the evolution of the system. The planet-planet interaction can be neglected since these quantities, which are very oscillatory, average out when the system has reached equilibrium.
		
		Figure~\ref{fig:torqpowcav} shows the torque~(top) and power~(bottom) acting on the planets as a function of time for a single planet model (1p10) on the left and the reference two-planet model (2p20) on the right. In these plots, the contribution of the indirect term has been taken into account in the calculation of the torque and power, as our coordinate system is centred on the star. Hence, the torque or power is the sum of the disc contribution, denoted by the subscript 'disc', and that of the indirect term, labelled 'ind'. The disc torque and the disc power are calculated as
		\begin{equation}
		 \label{eq:disctorq}
			\Gamma_{\rm disc} = \sum_{i,j=0}^{N_{\rm r}, N_{\rm \phi}} \vec{r}_{\rm p}\times \vec{F}_{{\rm p}ij},\\
			P_{\rm disc} = \sum_{i,j=0}^{N_{\rm r}, N_{\rm \phi}} \vec{r}_{\rm p} \cdot \vec{F}_{{\rm p}ij},
		\end{equation}
		where $\vec{r}_{\rm p}$ is the location of the planet and $\vec{F}_{{\rm p}ij}$ is the gravitational force between the planet and the mass in the $ij$th cell of the disc. Since the calculations have been performed in the star-centred non-inertial frame, there is an additional torque r power component due to the frame acceleration $\vec{a}_{\star}$, which is the acceleration on the star exerted by the disc. The indirect torque or power vanishes when the disc is completely symmetric and there is only one planet in the system. Otherwise, this term must be taken into account. The indirect torque $\Gamma_{\rm ind}$ and indirect power $P_{\rm ind}$ are calculated using the following equations
		\begin{equation}
		\label{eq:indtorq}
			\Gamma_{\rm ind} = M_{\rm p} \vec{r}_{\rm p}\times \vec{a}_{\star}, \\
			P_{\rm ind} = M_{\rm p}\vec{r}_{\rm p} \cdot \vec{a}_{\star}.
		\end{equation}
		
		\begin{figure}
			\centerline{\includegraphics[width=\columnwidth]{ 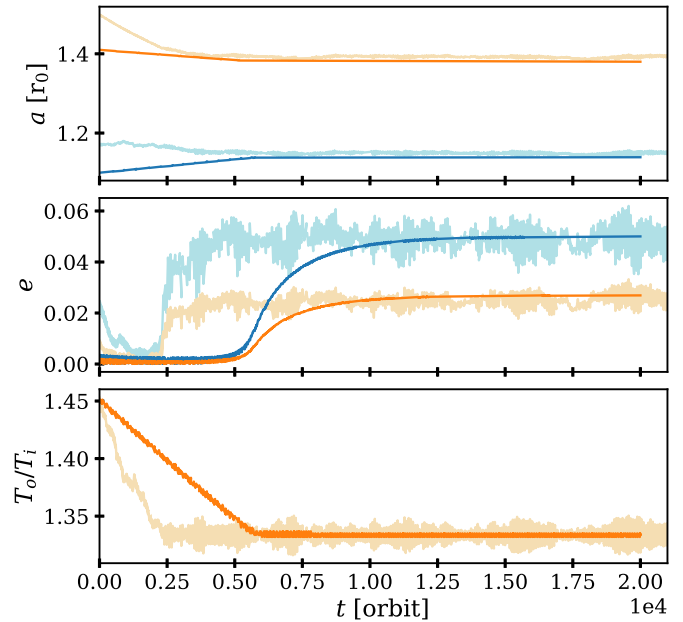}}
			\caption{Results of the N-body simulation with $\tau_{a_i}=1.6\times 10^5$, $\tau_{\rm a_o}=-2.7\times 10^5$, $\tau_{\rm e_i}=\tau_{\rm e_o}=-4\times 10^3$, and $M_{\rm i} = 10$, $M_{\rm o} = 20\, \rm M_{\oplus}$ are over-plotted on those of 2p20 \dz~ model. The dark and light colours refer to the N-body and hydrodynamical simulation. To facilitate the comparison, we shifted the time axis for the hydrodynamical simulation such that the $t=0$~orbit corresponds to the time at which the planets are just outside of 4:3 resonance.}
			\label{fig:nbodycav}
		\end{figure}
		
		The left column of Fig.~\ref{fig:torqpowcav} shows that the torque and power are zero for the single planet of 1p10 \dz~model. On the right, similar quantities are presented for the 2p20 \dz~model with two planets (left column in Fig.~\ref{fig:21020}). When the planets are trapped in their final halting positions, at $t \sim 10^{4}$~orbits, the torque and power acting on the inner planet are positive and those of the outer planet are negative, while their sum cancels out. When we consider each planet individually, we would expect the inner planet to move outwards because of the positive power, and the opposite for the outer planet. To maintain an equilibrium, the two powers need to be equal but have opposite sign. This results in zero net migration for the system as a whole. While the migration of the system is governed by the total power on the system, the vanishing total torque indicates that the eccentricities in the system no longer change either. When the calculation is carried out in the centre-of-mass frame, the sum of the power on all objects including the star should vanishe to maintain the equilibrium.

		To examine our claim in more detail, we performed additional N-body simulations using analogous parameters to the 2p20 \dz\ model. To calculate an equilibrium situation, we considered a system of two planets, in which the inner planet is migrating outwards and the outer planet inwards. Upon capture in 4:3 resonance (as observed in the hydrodynamical simulation), they should come to a halt and reach an equilibrium. When the migration of the planetary system stops, its energy and angular momentum remain constant on average. This implies that the total power should vanish, $P_\mathrm{i}+P_\mathrm{o}=0$. Here, we neglect the interaction energy between the two planets as it averages out after they reach their parking position at the inner edge of the disc. Using this equilibrium condition and the relation
		\begin{equation}
			\label{eq:power}
			P=|E| \, \frac{\dot{a}}{a} \, \equiv \, \frac{|E|}{\tau_{\rm a}} \,,
		\end{equation}
		which follows from Eq.~(\ref{eq:dotE}), we obtain the ratio of the migration rate of the inner to the outer planet as
		\begin{equation}
		\label{eq:mig-ratio}
			\frac{\tau_{\rm a_o}}{\tau_{\rm a_i}} = -\frac{a_{\rm i}}{a_{\rm o}} \frac{M_{\rm o}}{M_{\rm i}} \, 
			 = \, -\left(\frac{T_{\rm o}}{T_{\rm i}}\right)^{-2/3} \frac{M_{\rm o}}{M_{\rm i}}.
		\end{equation}
		Because our system is in 4:3 resonance, $T_{\rm o}/T_{\rm i}=4/3$, and the mass of outer planet is twice of the inner, we obtain $|\tau_{\rm a_o}/\tau_{\rm a_i}| \simeq 1.65$. Substituting the power and the final semi-major axis of the inner planet from Figs.~\ref{fig:torqpowcav} and \ref{fig:21020} in Eq.~(\ref{eq:power}), we obtain $\tau_{\rm a_i} \simeq 1.6 \times 10^{5}$. The migration of the outer planet is simply \num{1.65} times slower than that of the inner planet and {inwards}, meaning $\tau_{\rm a_o} \simeq -2.7 \times 10^5$. We input these timescales in an N-body code \citep{2005A&A...437..727K} that solves the equation of motion of the planets by the method of \cite{2002ApJ...567..596L}. To ensure that the system settles into the desired resonance, we placed the planets just outside of the 4:3 resonance. In addition to the migration timescales, eccentricity damping timescales $\tau_{\rm e_i}$ and $\tau_{\rm e_o}$ are also needed. For simplicity, we assumed that they are equal and find $\tau_{\rm e_i}=\tau_{\rm e_o}=-4\times 10^3$ to give a good match between the N-body and hydrodynamical results. From Eqs.~(\ref{eq:dotL}) and (\ref{eq:energy-angmom}) we have
		\begin{equation}
		\label{eq:tortaue}
			\frac{\Gamma}{L}  = \frac{1}{2} \frac{\dot{a}}{a}- \frac{e^2}{1-e^2} \frac{\dot{e}}{e} \, 
			\, \equiv \, \frac{1}{2} \frac{1}{\tau_{\rm a}}- \frac{e^2}{1-e^2} \frac{1}{\tau_{\rm e}} \,.
		\end{equation}
		With the obtained values for the damping timescales, the corresponding torques are $\Gamma_{\rm i} \simeq 1.18\times 10^{-10}$ and $\Gamma_{\rm o} \simeq -1.19\times 10^{-10}$. These values are indicated in the upper right panel of Fig.~\ref{fig:torqpowcav} with dashed lines, and they agree well with the averaged torque on the planets obtained from the hydrodynamical simulation.
		
		Figure~\ref{fig:nbodycav} demonstrates the results of the N-body compared to the final part of the hydrodynamical simulation. In the N-body simulation, the final resonance and locations match the hydrodynamical results, which implies that the proper ratio of migration timescales has been used, in agreement with Eq.~(\ref{eq:mig-ratio}). 
		For a different ratio, the planets would not reach equilibrium, but would migrate jointly, either outwards or inwards. We recall that the power and torque need to be calculated by including the indirect term when working in an accelerated reference frame. 
		
		From the equilibrium of the torque, $\Gamma_{\rm i} + \Gamma_{\rm o} = 0$, we can derive another important relation for the eccentricity damping timescales at the final parking position of the planet pair. Using Eq.\,(\ref{eq:energy-angmom}) for angular momentum and energy, Eq.~(\ref{eq:mig-ratio}) from power equilibrium, and assuming that $e^2 \ll 1$ (which is well fulfilled in our cases), we find 
		\begin{equation}
    	  \label{eq:tauecc-ratio}
		   \frac{ \left( 1 - 2 e_{\rm i}^2 \frac{\tau_{\rm a_i}}{\tau_{\rm e_i}} \right) }   
          	{ \left( 1 - 2 e_{\rm o}^2 \frac{\tau_{\rm a_o}}{\tau_{\rm e_o}} \right) }  \, = \, \frac{T_{\rm o}}{T_{\rm i}}  \,. 
		\end{equation}
		This implies that the final eccentricities in equilibrium are determined by the ratio of migration over eccentricity timescale $\tau_{\rm a}/\tau_{\rm e}$. We recall that $\tau_{\rm a_i}$ is positive and  $\tau_{\rm a_o}$ negative. The eccentricity damping timescales are always negative and much shorter that the migration timescales. Using the specified timescales and the equilibrium eccentricities of the N-body simulations, $e_{\rm i}=0.05$ and $e_{\rm o}=0.027$, we found that relation (\ref{eq:tauecc-ratio}) is fulfilled very well.

		\begin{figure}
			\centerline{\includegraphics[width=\columnwidth]{ 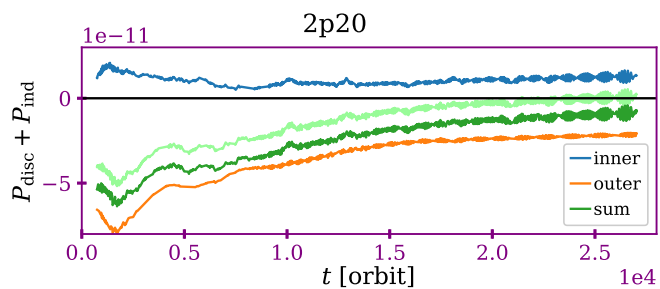}}
			\caption{Time evolution of the power on each planet and total power for the 2p20\,\ib~ model. The pale green line is the total power without the indirect term. The inward migration of the whole system can only be explained when the indirect term is considered in calculating the torques and powers.}
			\label{fig:torqpowsig}
		\end{figure}
		
		As another example, we show the power for the 2p20 \ib~model in Fig.~\ref{fig:torqpowsig}. This system migrates inwards after the resonance forms, but with a slower rate than that of the outer planet before the resonance. The total power on this system is negative, indicating inward migration, but its magnitude is lower than that of the outer planet. Without considering the indirect term, the total power on the system would be about zero, which is inconsistent with the inward migration of the system. This second example again demonstrates the necessity of including the indirect term, and shows that the total power on the system determines the migration of the resonant chain. 
		
		\begin{figure*}
			\centerline{\includegraphics[width=2\columnwidth]{ 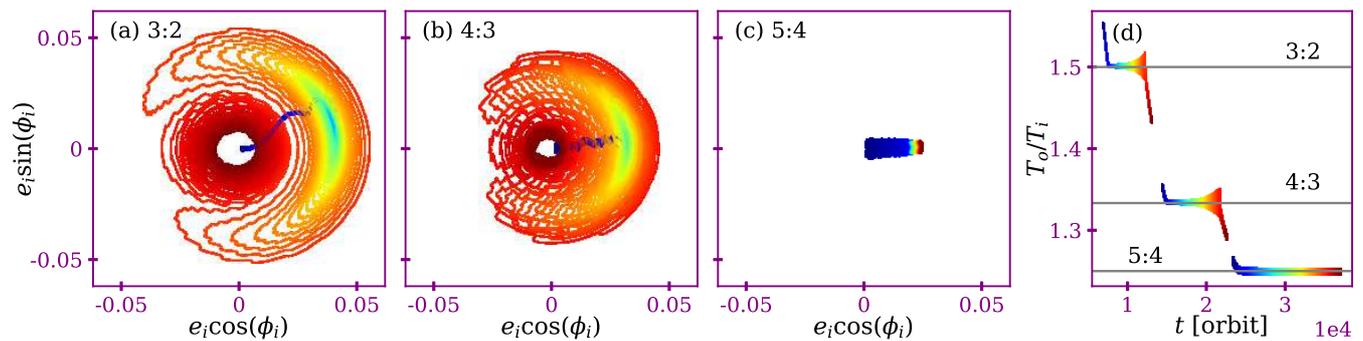}}
			\caption{Evolution of the eccentricity and resonant angle for the 2p20LH \ib~ model when the planets are engaged in 3:2, 4:3, and 5:4 resonances, respectively (panels~(a)--(c)). Panel~(d) shows the evolution of orbital period ratio. The colour-coding corresponds to the right panel and indicates the time evolution of the noted resonances in panels~(a)--(c). Explicitly, blue in panel~(a)--(c) indicates the time before the system forms resonance, which is noted at the top of each panel and also in panel~(d), and red represents the time after the system is in resonance. The exact time can be extracted from comparing the colours in panel~(d) with those in panel~(a)--(c).} 
			\label{fig:ephi}
		\end{figure*}

	\section{Comparing N-body to hydrodynamical simulations}
	\label{sec:nbodyvshydro}
		\begin{figure}
			\centerline{\includegraphics[width=\columnwidth]{ 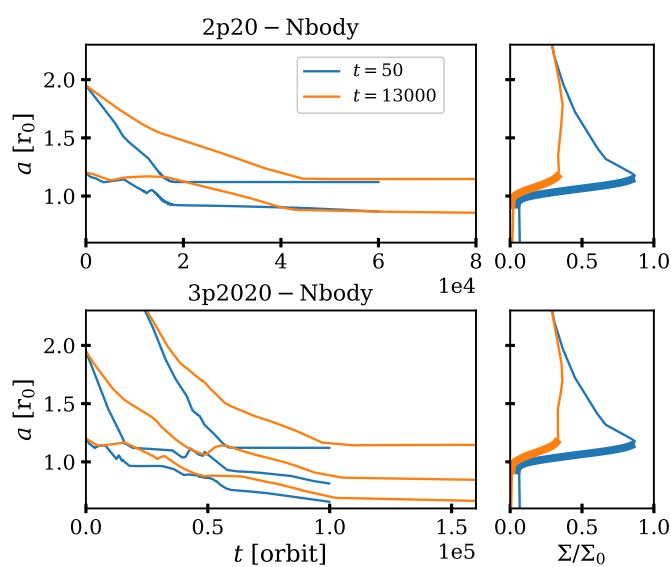}}
			\caption{Surface density in the horizontal axis and distance from the star on the vertical (right panel). The surface density profiles at $t=50$ and $t=\num{13000}$~orbit are denoted by blue and orange. The thick sections represent the areas in which the torque on a $10\, \rm M_{\oplus}$ circular planet is positive. Results of N-body simulations for the initial conditions as 2p20 (top left) and 3p2020 (bottom left) \dz\ models. The colours correspond to the surface density profiles shown in the right panels. The push by the outer planet in each pair can shove the inner planet out of the trapping point.}
			\label{figapp:nbody2p3p}
		\end{figure}
		
		In this section we compare the outcome of selected hydrodynamical models to customised N-body simulations, using the method described in Appendix~\ref{appendix:A}. When we adopt the density and temperature distributions from the hydrodynamical models, the N-body simulations  calculate the migration of the planets. As in our hydrodynamical simulations, we used $\epsilon = 0.6$ for the smoothing length of the gravitational potential of the planet (see Sect.~\ref{sec:modelling}). The temperature was constant, which translates into $\beta=0$. To calculate $\alpha_{\rm \Sigma}$, we took the surface density profile from the hydrodynamical simulation at two specific times. The viscosity transition was also taken into account as in Eq.~(\ref{eq:visc}). 
		
		In Fig.~\ref{figapp:nbody2p3p} we display the results of two sets of N-body simulations using the initial conditions of the 2p20 and 3p2020 \dz\ models, where the mass of inner planet is $10\, \rm M_{\oplus}$ and that of the outer planet(s) $20\, \rm M_{\oplus}$. Each set contained two identical setups and differed only in their surface density profile, which was chosen at $t=50$~orbit (unperturbed) and $t=\num{13000}$~orbit (evolved). We chose two snapshots to take a possible change of the surface density in the hydrodynamical simulations into account. The left column in this figure contains the migration of the planets, and on the right the plot the surface density profiles. The thicker sections in the right panels mark the region in which the torque is positive. The upper and lower rows show the results for the model with two and three planets. In all of these models, the inner planets in each model is pushed by the outer planets into the positive torque region regardless of which surface density profile is used. The only difference between the outcomes of these simulations is the slower migration of the planets in the model with the evolved surface density profile. This slower migration is caused by the flattened bump that is created by the planets in the hydrodynamical simulations. 
		
		In the two-planet simulations, the inner planet is first trapped at the zero-torque location and then pushed farther in by the incoming (more massive) outer planet. The migration of the outer planet is halted when it reaches the trap at the outer edge of the viscosity transition. At this time, the inner planet has been pushed into the negative torque region again and slowly migrates inwards. However, in the three-planet cases, the two inner planets are first trapped in a resonant configuration while the outer planet approaches. Later, at $t \approx 50\,000$, the outer planet catches the inner pair in resonance and pushes them out of the positive torque region into the inner disc. The migration of the inner planets can only be halted if they are still in the positive torque region when the outer planet reaches its final trapping location. In all of our N-body cases, we saw that the outermost planet was able to push the inner planets inwards beyond the trap region into to the negative torque region, such that they continue a very slow inward migration. In strong contrast, in the corresponding hydrodynamical simulations, the inner planet was able to stop the inward migration of the planetary system, even though an unstable evolution occurred in the three-planet case. 
		
		\begin{figure}
			\centerline{\includegraphics[width=\columnwidth]{ 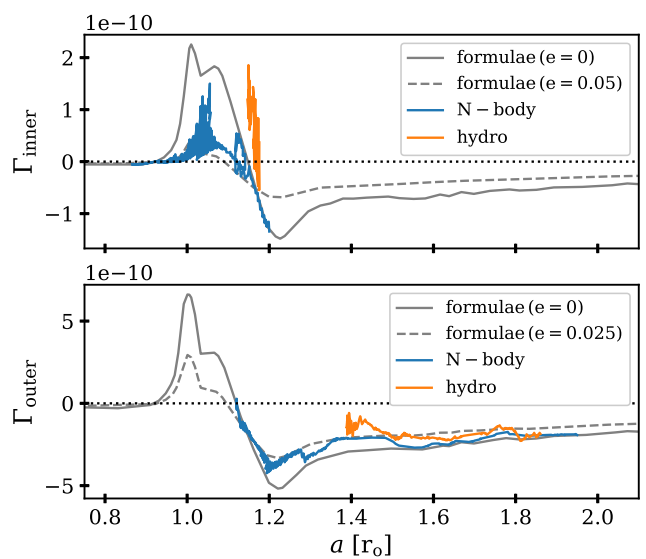}}
			\caption{Torque from the hydrodynamical and N-body simulations vs. the semi-major axes of the planet for the inner (top) and outer planet (bottom) in 2p20 \dz\ model. The grey lines denote the torque calculated from the formulae used in the N-body code as a function of distance from the star for a circular (solid) and an eccentric (dashed) planet. The values of the used eccentricities equal those of the hydrodynamical simulation after equilibrium. Comparison of the grey lines and the outputs of the simulations shows that these formulae are capable of correctly calculating the torque only for the outer planet before it approaches the transition zone.}
			\label{figapp:torques}
		\end{figure}
		
		Our results indicate that the strong positive torques experienced by the planets at the trap in the hydrodynamical cases are not fully reflected by the N-body simulations. This is depicted in Fig.~\ref{figapp:torques}, which shows the torque from the hydrodynamical and N-body simulations (using the unperturbed surface density profile) against the semi-major axes of the planets. 
		The grey solid and dashed lines are torques calculated using the formulae in the N-body code for a circular and an eccentric planet. For the outer planet (lower panel), the torque agrees excellently between the hydrodynamical and N-body models down to $a \sim 1.5$. The discrepancy increases when the transition edge is approached. Especially in the hydrodynamical model, the torque on the inner planet (upper panel of Fig.~\ref{figapp:torques}) is positive and relatively strong compared to the value given by the formulae. Therefore it can compensate for the push by the outer planet and consequently halts at $r \sim 1.4$. In the N-body simulation, the torque on the inner planet is mostly negative. Even when the torque is positive inside the transition zone, it is much weaker than that of the hydrodynamical simulation. This indicates that the actual hydrodynamical values for torque and eccentricity damping ($\tau_{\rm L}, \tau_{\rm e}$) inside and close to a transition zone differ from those calculated from the formulae that are estimated from relatively smooth disc surface density profiles. We would like to mention that the N-body torque would be even weaker if the evolved surface density profile were used in Fig.~\ref{figapp:torques} because both the lower surface density and the flatter profile would result in weaker torques. 
		
\begin{figure}
			\centerline{\includegraphics[width=\columnwidth]{ 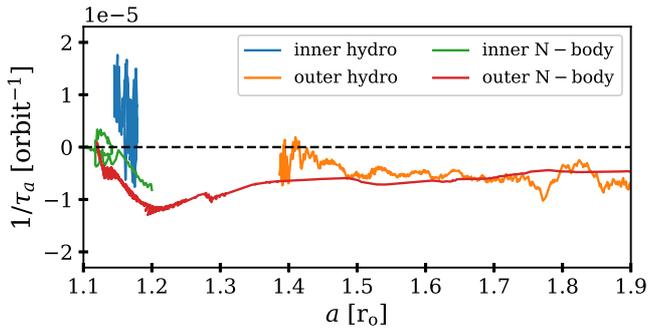}}
			\caption{Inverse of the migration timescale for the hydrodynamical and N-body simulation of the 2p20 \dz\ model.}
			\label{figapp:invtaua}
		\end{figure}

		For illustrative purposes and to connect to the next example, we show the inverse of the migration timescales. Because the migration timescale is extremely large for the trapped planets, showing its inverse is more suitable for comparison.  for the hydrodynamical and N-body simulation in Fig.~\ref{figapp:invtaua} . The migration timescale of the inner planet in the hydrodynamical simulation is much smaller than  its N-body counterpart. For the outer planet, the migration timescale is very similar in both simulations until $a\sim1.5$. As the planet approaches the transition zone, the migration timescale in the hydrodynamical model deviates from that of the N-body.
		
		To verify our suggestion that the difference between hydrodynamical and N-body simulations hinges n a very weak positive torque near and inside of the trap region in the N-body runs, we performed a set of additional N-body simulations in which the migration timescale was given by an analytical model. For each planet, we chose a negative constant reference migration time, $\tau_{\rm a_0}$, except inside a transition zone, where it became positive in order to mimic the planet trap. Mathematically speaking, the migration timescale is given by $\tau_{\rm a} = \tau_{\rm a_0} /f_{\rm trans}$ with $f_{\rm trans}$ defined as
		\begin{align}
			\label{eqapp:ftrans}
			f_{\rm trans} &= 1+ a \left[ \tanh\left(\frac{r-r_{\rm ref}}{\Delta}\right) -1 \right] + \\ \nonumber
			&+ b\, \left(\frac{r-(1+c \Delta)r_{\rm ref}}{\Delta}\right)\, 
			\exp \left(- \left(\frac{r-r_{\rm ref}}{\Delta}\right)^2 \right)  \, ,
		\end{align}
		where $r_{\rm ref}$ is the location of the transition, $\Delta$ determines the width, and $a$, $b$, and $c$ control the shape of the transition zone: $a$ is the difference between inner and outer inward migration, $b$ is the strength of the trap, and $c$ is the bump strength. The analytical form of this function is motivated by the results from our hydrodynamical modelling. Different sets of these parameters correspond to different shapes for the disc inner edge. For example, the third term adjusts the torque if a surface density bump exists before the zero-torque location. We causion that finding a reverse process that gives the exact shape of the disc inner edge is not trivial because of the torque calculation complexities. It is very hard if not impossible to find the shape and the disc properties only by having the torque on the planet. Equation~\ref{eqapp:ftrans} is a general profile that can be used for any type of planetary trap only by proper adjustment of the paramters, either the trap is a dead-zone or disc inner edge. However, we note that in the \ib~ mode, the paramters would be time dependent becaue of the change in the surface density profile with time.
		
		This function is defined such that it gives a faster migration in the outer than the inner disc plus a region where the migration becomes very slow and outwards. Far from $r_{\rm ref}$, where $f_{\rm trans}=1$, the migration of the planet is simply inwards with the timescale of $\tau_{\rm a_0}$. As the planet approaches $r_{\rm ref}$, $f_{\rm trans}$ becomes initially larger than unity. This is where the migration of the planet becomes faster because of the surface density bump before the disc edge. Afterwards, $f_{\rm trans}$ decreases until it becomes zero and brings the migration of the planet to a halt. If by some means, for example the push of another planet, the planet moves to where $r <r_{\rm ref}$, a negative $f_{\rm trans}$ causes an outward migration similar to a disc inner edge. If the planet is pushed even more inwards, such than the planet is out of the disc inner edge, we expect that the migration of the planet again turns inwards, but very slowly becuase the surface density is low. This is what $f_{\rm trans}$ does for $r \ll r_{\rm ref}$. We would like to insist that this factor does not produce the torque from the asymmetric features such as vortices, which produce usually a stochastic torque. 
		
\begin{figure}
			\centerline{\includegraphics[width=\columnwidth]{ 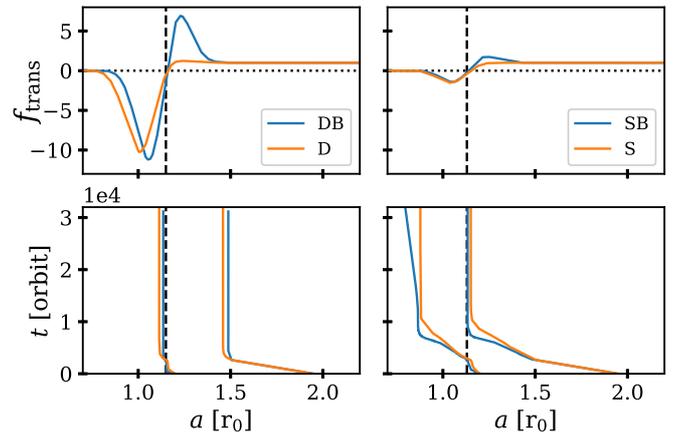}}
			\caption{Results of four N-body simulations in which a migration timescale transition zone, as depicted with the same colour as in the top panels, is used. The vertical dashed line corresponds to $f_{\rm trans}=0$, and marks the transition from positive to negative migration.}
			\label{figapp:arttran}
		\end{figure}

		We examined two sets of models, one with a deep and one with a shallow drop in $f_{\rm trans}$, resembling the hydrodynamical and N-body simulations, respectively. In each set, one has a bump outside of the drop and one does not. The latter models were performed in order to confirm the role of the bump, which  speeds up the inward migration just outside of the transition zone, as seen in the hydrodynamical models. The upper panels of Fig.~\ref{figapp:arttran} show $f_{\rm trans}$ for the deep (left) and shallow (right) drop models. The parameters we used in these profiles are listed in table~\ref{tab:ftranspara}.
		
		\begin{table}
			\centering
			\begin{tabular}{ccccc}
				\toprule[2pt]
				Parameter & DB & D & SB & S \\
				\midrule[1.5pt]
				$\Delta$ & 0.12 & 0.12 & 0.12 & 0.12 \\
				$r_{\rm ref}$ & 1.13 & 1.05 & 1.13 & 1.09 \\
				$a$ & 0.45 & 0.45 & 0.45 & 0.45 \\
				$b$ & 20 & 8 & 3 & 1.4 \\
				$c$ & 0.2 & 1.1 & 0.2 & 1.1\\
				\bottomrule
			\end{tabular}
			\caption{List of parameters for $f_{\rm trans}$ profiles shown in the top row of Fig.~\ref{figapp:arttran}.
				The letters are D for the deep trap, S for the shallow trap, and B for the bump.}
			\label{tab:ftranspara}
		\end{table}
		
		We used the same masses and initial locations as 2p20 and considered $\tau_{\rm a_o}=-10^{4}$ and $\tau_{\rm a_i}=-2\times10^{4}$. The eccentricity timescales were fixed to  $\tau_{\rm e_i}=\tau_{\rm e_o}=-3\times10^{3}$. These numbers are similar to those of Sec.~\ref{sec:torpow}.  The result of the simulations is shown in the bottom panels of Fig.~\ref{figapp:arttran} with the corresponding colours for each of the transition profiles. 
		
		In all of the models, the planets form 3:2 resonance after about $\num{2800}$~orbits. In the deep models (left panel), the migration of the planets is halted when the total power acting on both planets vanishes. This location is near $f_{\rm trans}=0$, which is denoted by a vertical dashed line. There is no significant difference between the models with and without a bump. In the shallow model with a bump, the inner planet continues its migration to the transition zone and is not able to stop the migration of the outer planet. Finally, the inner planet leaves the transition region while the outer planet remains at the trapping location, and therefore the resonance breaks. However, in the model without a bump, the migration of the inner planet halts when the outer planet reaches the trapping point and they remain in 3:2 resonance. This different behaviour arises because in spite of the S model, the inner planet has passed the transition zone before the outer planet approaches the trapping point in the SB model. 
		
		To summarise, based on the results in this section, we found that the N-body simulations favour resonant pushing, unlike hydrodynamical simulations. The reason might be that the formulae that are used in the N-body calculations do not give proper values where the surface density deviates from the simple power-law profiles. This point requires more detailed analyses in a separate study.

	\section{Over-stability}
	\label{sec:overstability}
		
		In the results of our simulations in Sec.~\ref{sec:results}, the \ib~models with lower aspect ratio, $h_{0}=0.03$, show over-stable behaviour (see runs 2p20LH, 3p2020LH, and one-by-oneLH, as displayed in Figs.~\ref{fig:21020LH}, \ref{fig:3102020LH} and \ref{fig:oneoneLHsig}). Over-stability for planets at the inner edge of a protoplanetary disc was observed in N-body simulations by \cite{2015MNRAS.449.3043X}, but has not yet been reported convincingly in hydrodynamical simulations. \citet{2018MNRAS.474.3998H} attributed the mean-motion resonance breaking that occured in their main hydrodynamical simulations to over-stability.  However, the eccentricity evolution of the planets in their models is caused by crossing of the planets over the 2:1 resonance.
Only one model, that is presented in their appendix, possibly resembles overstable behavior.
		
		\begin{figure}
			\centerline{\includegraphics[width=\columnwidth]{ 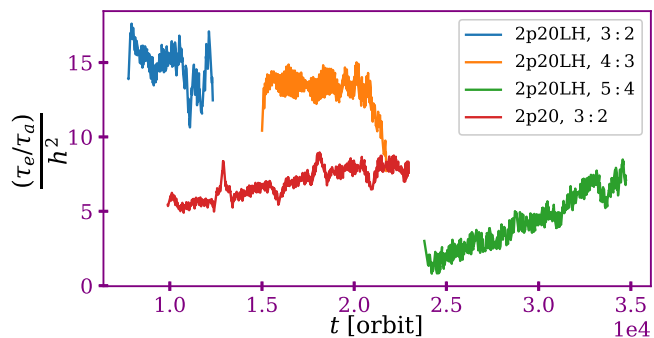}}
			\caption{Ratio of migration and eccentricity damping timescales scaled by the aspect ratio squared vs. time.}
			\label{fig:taue-taua}
		\end{figure}
				
		In an over-stable system, the planets successively enter various first-order resonances that are broken after a short time. In the transient resonances, the eccentricities oscillate around an equilibrium value with increasing amplitudes until the resonance breaks and the planets enter an even tighter resonance. Over-stability occurs when the eccentricity damping is not efficient enough. As a result, the eccentricities of the planets, that are excited due to their mutual gravitational interaction, gradually grow while the libration amplitude increases \citep{2014AJ....147...32G}. This growth continues until the resonance is eventually broken. 

		Figure~\ref{fig:ephi} shows the over-stable behaviour in more detail for the 2p20LH \ib~ model, in which the system becomes over-stable during the 3:2 and 4:3 resonances (see also Fig.~\ref{fig:21020LH}). Panels~(a),~(b), and~(c) demonstrate the evolution of the eccentricity $e$ and the resonant angle $\phi$ of the inner planet during the 3:2, 4:3, and 5:4 resonances. The distance of every point in these panels from the centre $(0,0)$ represents the eccentricity, and its angle with the horizontal axis is a measure of the resonant angle. The colour scheme indicates the time evolution as shown in panel~(d), where the time evolution of the orbital period ratio is plotted. The resonant angle for a first-order $j+1:j$ resonance is defined as $\phi_{\rm i}=(j+1)\lambda_{\rm o} - j \lambda_{\rm i} - \varpi_{\rm i}$, where $\lambda$ and $\varpi$ denote the mean longitude and longitude of the pericentre, respectively. If $\phi_{\rm i}$, which measures the angle between the conjunction and the pericentre of the inner planet, librates around a constant value, the system is in a resonance. Panel~(a) implies that as the planets settle into the 3:2 resonance, the eccentricity of the inner planet grows to $\sim 0.04$ and the resonant angle librates around \ang{3}. The amplitude of the libration as well as the eccentricity increases until it circulates and the system leaves the resonance. The situation for the 4:3 resonance in panel~(b) is very similar, except that the equilibrium eccentricity is lower and the resonant angle is smaller. In contrast, the final resonance 5:4 appears to be stable around $e\sim 0.025$ and $\phi_{\rm i}=0$.
		
		 The stability of first-order mean motion resonances between planets assuming constant eccentricity damping and migration rates was studied by \cite{2014AJ....147...32G}. They showed that these resonance configurations are stable if $\tau_{\rm e}/\tau_{\rm a}$ is small enough. They also presented a condition between the equilibrium eccentricity $e_{\rm eq}$ and mass of the planets. They suggested that if $e_{\rm eq} \lesssim ({\rm M_{\rm p}/M_*})^{1/3}$, the system remains stable. For the inner planet in our 2p20LH \ib~ model, $({\rm M_{\rm p}/M_*})^{1/3} \sim 0.03$. Therefore their stability criterion is only satisfied in our case for the 5:4 resonance. They also found that $e_{\rm eq} \sim (\tau_{\rm e}/\tau_{\rm a})^{1/2} \sim h$ that predicts $e_{\rm eq} \sim 0.03$, which roughly agrees with our results. However, \cite{2018MNRAS.481.1538X} used a migration model based on planet-disc interaction studies, suggests that the equilibrium eccentricity of the resonant planets is higher than those given by the studies that used constant migration and eccentricity damping rates. Our results seem to match \cite{2014AJ....147...32G} better. 
		
    	It has been shown that $\tau_{\rm e}/\tau_{\rm a}$ is of the order of $h^2$ \citep[e.g.][]{2004ApJ...602..388T}. Therefore we expect that over-stability more probable occurs in thicker discs, while we observe it for our thinner models. This may be explained by noting that the planets in the models with $h_{0}=0.03$ open a partial gap around their orbits. Although this partial emptiness around the orbit of the planets slows their migration down, it also decreases the efficiency of the eccentricity damping. Therefore our simulations apparently show that the net effect of the gap opening is more in the favour of larger $\tau_{\rm e}/\tau_{\rm a}$. Figure~\ref{fig:taue-taua} shows $\tau_{\rm e}/\tau_{\rm a}$ scaled by $h^2$ for model 2p20LH and 2p20 for when the planets are in a resonance. In the 2020LH model, $\tau_{\rm e}/\tau_{\rm a}$ is larger at the over-stable resonances than in the stable resonance. Interestingly, the scaled ratio has similar values for the stable resonance in the 2p20 and 2p20LH models. However, this speculation needs to be justified by more rigorous studies. 
		
	\section{Stability of the systems and comparison with observation}
	\label{sec:stability}
\begin{figure}
			\centerline{\includegraphics[width=\columnwidth]{ 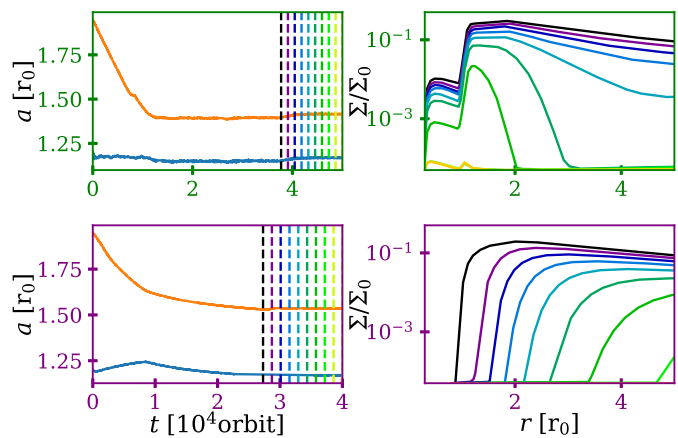}}
			\caption{Time evolution of models 2p20 after photo-evaporation is activated. The left panels show the semi-major axes of the planets, and the right panels demonstrate the surface density evolution of the disc. The vertical lines in the left panels correspond to the time when the surface densities are plotted.}
			\label{fig:photoevap}
		\end{figure}

		Formation of planetary systems is a long journey from when they form in a protoplanetary disc until the disc dispersal. It is therefore uncertain whether those of our systems that survived instability during the simulation time could remain stable afterwards. Because continuing all of our simulations would be computationally very costly, we only examined whether the \dz~and \ib~2p20 and 2p20LH models retained their configuration during the disc dispersal. We included the X-ray photo-evaporation model of \cite{2012MNRAS.422.1880O}  with an X-ray luminosity of $L_{\rm x} = 10^{32} \rm erg/s$ and continued these simulations for several thousands orbits until the disc around the planets was dispersed. This high value of $L_{\rm x}$ was chosen to speed up the simulations. The outcome of these tests shows that none of them become unstable (Fig.~\ref{fig:photoevap} and \ref{fig:photoevap03}). Similar outocmes can be seen in the results of the N-body simulations that include photo-evaporation \citep[e.g.][]{2015MNRAS.453.1632M,2016MNRAS.458.2051M}.

		To inspect the stability of other models, we used the criteria presented by \cite{1996Icar..119..261C} and \cite{1993Icar..106..247G}, who suggested that a low-mass planetary system would remain stable if $\Delta/R_{\rm H} > 2 \sqrt{3}$ for two-planet systems, and would always be unstable if $\Delta/R_{\rm H} < 10$ for more than two planets. Here, $\Delta$ is the initial difference of the semi-major axes of the planets, and $R_{\rm H}$ is their mutual Hill radius. For all of our two-planet models, this condition is well satisfied. Only \ib~3p1020 of the multi-planet models has $\Delta/R_{\rm H}$ higher than 10. This means that only  the two-planet system would survive after a long-term evolution.
		
		We also compared these stable systems with observed exoplanets\footnote{\url{http://www.openexoplanetcatalogue.com/}} and found two systems, Kepler-804 and K2-189, with configurations close to our results. Kepler-804 has two planets revolving around a solar-type star \citep{2016ApJ...822...86M}. The size of the outer planet is about twice that of the inner one (i.e. almost an Earth-size planet) and has an outer-to-inner period ratio of $\sim 14.3/9.6 \simeq 1.5$. Using the mass-radius relation in \cite{2016ApJ...825...19W} for sub-Neptune-sized planets, the mass and period ratios of this system resemble those in our \dz~2p20 model (although the masses are lower than ours). The other system, K2-189, also has two planets with radii of about 2.48 and 1.51 times the radius of Earth and revolve around a star with the mass of $0.93 M_{\odot}$  \cite{2018AJ....155..136M}. The period ratio of the planets is about 1.3, which is near the 4:3 commensurability. The relation in \cite{2016ApJ...825...19W} gives a mass ratio of the outer to inner planet of $\sim 1.9$. These mass and orbital period ratios are similar to those in our \ib~2p20 model. Therefore both of our tested inner edge models could be used to explain the formation of these observed close-in exoplanetary systems.
	
	\section{Summary and conclusion}
	\label{sec:summ}
		Using hydrodynamical simulations, we investigated whether a trapped planet at a \deadzone\ (\dz) or a \innerbou\ (\ib) can be pushed into the inner cavity towards the central star by a resonant chain. We examined systems with two and three planets in which the planets were simultaneously placed in the disc, and systems with more than three planets in which the planets were added one by one after the previous planets formed a resonant chain. 
		The \dz\ model has a very narrow and steep transition in the surface density, and the \ib\ model has a much shallower profile. The mass of the inner planet was always $10\, \rm M_{\oplus}$ while the outer planets had either the same mass or were more massive. We summarise and discuss our findings below. 
		
		The steepness of the inner disc edge, that is, the gradient of the surface density plays a key role in the dynamical evolution of our systems. The \dz~model, where the sign of the power (and torque) abruptly changes at the edge of the transition zone, in which the planet experiences positive power, acts as a dam that can stop an incoming resonant chain of planets. The inner planet cannot pass this dam unless an instability in the system hurls it into the inner disc.  On the other hand, in the \ib\ model, which has a smoother transition and a fainter surface density profile, a resonant chain can push the inner planet into the inner cavity. However, even in this case, it requires multiple and/or more massive planets to push the inner planet across the trap. Test calculations using a broader transition zone in the \dz\ setups produce wider resonant systems that are less prone to instability These tests indicate that if the viscosity transition zone is very wide, the inner planet might be pushed out of its trapping point, similar to the \ib~models.
		
		In the two-planet cases, the final configurations in the \dz\ models are usually tighter than in their \ib\ counterparts. A typical outcome for a $10$ and $20\, \rm M_{\oplus}$ combination is the 4:3 resonance. Most of the three-planet systems in the \dz\ models became unstable and a scattering event between the two inner planets occurred, while in the \ib\ models, the planets were trapped in a 3:2:1 resonant chain. In the models where we added the planets one by one, increasing the number of planets has no effect on the innermost planet and only increases the possibility of instability.
			
		Planets in the models with $h_{0} = 0.03$ behave differently than those with $h_0=0.05$. In the cooler discs, the planets open partial gaps around their orbits. The gap opening changes both the migration rate and eccentricity damping such that over-stable oscillations of the planets' eccentricities occur \citep{2014AJ....147...32G} and the planets reach successively tighter resonances, jumping for example from 3:2 to 4:3 to 5:4. Our results are the first hydrodynamical simulations that clearly show the over-stable behaviour in resonant planetary systems, see Fig.~\ref{fig:ephi}.  For computational reasons we focussed our study on transitions located at 1\,au with typical aspect ratio of $h=0.05$. However, the inner edge of discs lies about ten times closer to the star where the disc is  much thinner $h = 0.02$ to $0.03$, and we expect that under these conditions, over-stability may occur frequently. This raises and highlights the need for more detailed studies of this topic. 
		
		In addition to the aspect ratio, we varied the disc mass (surface density, $\Sigma_0$) and viscosity ($\alpha$). Using a higher surface density for the disc only speeds up the evolution and has no specific effect on the final results. This is in agreement with early studies of resonant capture by \citet{2002ApJ...567..596L}, who showed that the ratio of migration over eccentricity damping determines the outcome of capture, and not the absolute timescale.
		Although the planets in the less viscous disc were able to slightly modify the surface density profile of the disc through partial gap opening, the outcomes of the models are not significantly different from those with higher viscosity.
			
		In the \dz\ simulations the steep positive surface density profile gives rise to the creation of vortices that might interfere in the migration of the planets. Our results show that the presence or number of vortices at the edge of a dead-zone does not influence the evolution of the system greatly, besides a small enhancement of the susceptibility to unstable evolutions.
			
		The eccentricity of the planets remain small as long as no instability happens or they are not pushed into the inner disc, where the surface density is very low. This indicates that the eccentricity damping timescales in our models are still much shorter than the migration timescales. The comparison of our hydrodynamical and N-body simulations indeed show a ratio of $|\tau_a/\tau_e| \approx 40$ for the inner $10\, \rm M_{\oplus}$ planet. 

		We analysed the conditions for an equilibrium parking of a system of two planets at the inner edge in detail. In agreement with previous results for migrating planets \citep{2007A&A...473..329C}, we showed that it is determined by a vanishing total power, for which all objects have to be taken into account. This could be formulated as a condition for the ratio of migration timescales, see Eq.~(\ref{eq:mig-ratio}). The additional requirement of vanishing total torque yields a condition for the equilibrium eccentricities, see Eq.~(\ref{eq:tauecc-ratio}). This last equation in particular implies that resonant planetary systems created by trapping must have a non-zero eccentricity. The validity of these relations was confirmed in our hydrodynamical and N-body simulations.

		As hydrodynamical simulations are very time-consuming, we performed additional customised N-body simulations in order to test the frequently used approach of taking analytical formulae for the planetary torque and eccentricity damping. In the appendix, we demonstrate that the direct usage of the disc parameters, namely density and temperature profiles, from the hydrodynamical simulations as inputs for the forces in the N-body formulae does result in different evolutions of the planetary systems. In particular, the inner planet in the N-body simulations was not able to stop the resonant chain. This indicates that the positive torque (power) contribution from the inner edge of the disc is underestimated by the analytical formulae. 

		The main conclusion of our study is that it is difficult for the planets to be pushed into the cavity created by a \dz\ through resonant migration due to its steep density slope. However, for a smoother planetary trap, such as an \ib\, a resonant chain can help push the planets inwards. When we consider a \dz~ that does not coincide with the inner boundary of the disc but lies farther out, the formed planets inside the dead-zone would have difficulties to reach the inner disc unless the resonant chain becomes unstable. Nonetheless, when they can pass the dam of the viscosity transition, they can easily move towards the disc inner edge by the help of resonant migration. 
		
		In order to compare our results with the observed exoplanetary system, we also examined the stability of our systems and found that it is more probably for the two-planetary systems to remain stable. We found that only two observed systems of Kepler-804 and K2-189 are similar to the outcome of our 2p20 models. Although, we cannot draw a conclusion based on these two observed systems, we suggest that the more packed systems are more likely to form in a disc with a steeper inner edge.
		
		We would like to note that our results were obtained using locally isothermal disc. The study by \cite{2016A&A...586A.105F}, who investigated planet trapping at a \dz~ in a non-isothermal disc using 2D and 3D magnetohydrodynamical simulations, indicated that the inner edge of a dead-zone can be a mass-dependent barrier for the migration of vortices that formed inside the dead-zone, and also planets. 
	
		As mentioned before, for computational reasons, we placed the inner boundary at a distance of 1\,au using typical values for the disc surface density, aspect ratio, and viscosity at that location.
		The models with lower aspect ratio or higher surface density may resemble the conditions closer to the star more. In particular, the thinner disc models ($h_{0}=0.03$) give an insight into how the resonant pushing can be different in the inner disc in the later stages of the disc evolution. We might therefore expect that over-stable librations with subsequent tightening of the resonant chains are a frequent outcome and may help in producing very compact resonant systems such as Kepler-223 \citep{2016Natur.533..509M}.
				
		This study has been carried out using a background disc profile that dictates divergent migration in the absence of a planetary trap. If a different disc setup were used that helps the converging migration, instability would be more probable.
		
		One of the remaining open questions is that what happens to the resonant planets once they are pushed into the inner cavity. In some \ib\ models, where the inner planet is pushed all the way to the region with very low surface density, its eccentricity increased to high values. Whether the system will remain stable afterwards is a question we were unable to answer using our hydrodynamical simulations.
		
		The tidal interaction between the star and the planets can be an important phenomenon for the planets that are located very close to the star. In this work, as the first hydrodynamical study that tries to simulate multiple planets at the disc inner edge, we ignored this effect for simplicity. How much the evolution can be changed by this effect is a question for future works.
			
	\begin{acknowledgements}
		We acknowledge the support of the DFG priority program SPP 1992 "Exploring the Diversity of Extrasolar Planets under grant KL 650/27". The authors also acknowledge support by the state of Baden-Württemberg through bwHPC. We also acknowledge the stimulating discussion with B.~Bitsch and D.~Carrera. We also thank the anonymous referee for her/his constructive comments.
	\end{acknowledgements}

	\bibliographystyle{aa}
	\bibliography{discedge}
	
	\begin{appendix}
\section{N-body simulations. Methods}
		\label{appendix:A}
		N-body simulations of migrating planets are a common alternative for the lengthy and computationally expensive hydrodynamical simulations. In studies where the long-term evolutions of many planetary systems are desired, using a hydrodynamical simulation is not only expensive but can even be not feasible. Using N-body simulations augmented with a suitable migration treatment therefore is a good option \citep[e.g.][]{2000MNRAS.315..823P,2009ApJ...699..824O,2014MNRAS.445..479C,2017MNRAS.470.1750I}. The basic ingredients for performing such a simulation is an N-body code, an expression for the disc torque acting on the planet, a relation for calculating the eccentricity damping timescale, and an equation of motion that modifies the acceleration of the planets accordingly. 
		
		In order to examine how well the results of such simulations agree with their hydrodynamical counterparts, we employed the \texttt{REBOUND} code\footnote{The N-body code \texttt{REBOUND} is freely available at http://github.com/hannorein/rebound.} \citep{2012A&A...537A.128R} with the WHFAST integrator \citep{2015MNRAS.452..376R} with a time step of $dt=10^{-3}$~orbits, and implemented the formulae of \cite{2010MNRAS.401.1950P,2011MNRAS.410..293P} into it to calculate the torques on the planets. The torque has two components: Lindblad, $\Gamma_{\rm L}$, and corotation, $\Gamma_{\rm C}$. These are both altered if the planet is eccentric. The corresponding correction factors for the Lindblad and corotation torques, $\Delta_{\rm L}$ and $\Delta_{\rm C}$, respectively, are taken from \cite{2008A&A...482..677C} and \cite{2014MNRAS.437...96F}. The formulae we used are identical to those employed by \cite{2014MNRAS.445..479C} and \cite{2017MNRAS.470.1750I}, but (a)~our simulations are 2D and therefore the inclination was set to zero, (b)~we applied the additional correction factors for the smoothing length of the planet as in \cite{2010MNRAS.401.1950P} for consistency, and (c)~because the disc is locally isothermal, we considered the thermal diffusivity $\chi$ as infinite and set the adiabatic index to $\gamma =1$. For completeness and reference, we briefly summarize the relevant formulae. The total torque is given by
		\begin{equation}
			\label{eq:totaltorque}
			\Gamma = \Delta_{\rm L} \Gamma_{\rm L} + \Delta_{\rm C} \Gamma_{\rm C}\,,
		\end{equation}
		where $\Gamma_{\rm C}$ is the sum of the two components, barotropic and entropy-related co-rotation torque, and each of them has a linear torque and a horseshoe drag \citep[see][]{2011MNRAS.410..293P}. Therefore, the co-rotation torque reads
		\begin{align}
			\label{eq:cotorque}
			&\Gamma_{\rm C} = \Gamma_{\rm c,baro} + \Gamma_{\rm c,ent}\,, \\
			&\Gamma_{\rm c,baro}= \Gamma_{\rm c,hs,baro} F(p_{\rm \nu}) G(p_{\rm \nu})+ \Gamma_{\rm c,lin,baro} (1-K(p_{\rm \nu}))\,,\\
			&\Gamma_{\rm c,ent}= \Gamma_{\rm c,lin,ent} \sqrt{(1-K(p_{\rm \nu}))}\,.
		\end{align}
		The functions $F(p_{\rm \nu})$, $G(p_{\rm \nu})$, and $K(p_{\rm \nu})$ regulate the saturation of the co-rotation torque due to the disc viscosity, where $p_{\rm \nu}$ is defined as $2/3 \sqrt{a^2 \Omega x_{\rm s}^3/2\pi \nu}$, with $x_{\rm s}$ being the half-horseshoe width of the planet. For brevity, we avoid listing the definition of the functions and refer to  \cite{2011MNRAS.410..293P} or \cite{2017MNRAS.470.1750I}. The entropy-related horseshoe drag vanishes due to the locally isothermal assumption. The Lindblad torque and the components of the co-rotation torque are as follows:
		\begin{align}
			\label{eq:torquecomponents}
			&\frac{\Gamma_{\rm L}}{\Gamma_{0}} = (-2.5-1.7\beta + 0.1\alpha_{\rm \Sigma})\left(\frac{0.4}{\epsilon}\right)^{0.71}\,,\\
			&\frac{\Gamma_{\rm c,hs,baro}}{\Gamma_{0}} = 1.1(1.5-\alpha_{\rm \Sigma}) \left(\frac{0.4}{\epsilon}\right)\,,\\
			&\frac{\Gamma_{\rm c,lin,baro}}{\Gamma_{0}} = 0.7(1.5-\alpha_{\rm \Sigma}) \left(\frac{0.4}{\epsilon}\right)^{1.26}\,,\\
			&\frac{\Gamma_{\rm c,lin,ent}}{\Gamma_{0}} = 2.2\beta \left(\frac{0.4}{\epsilon}\right)^{0.71} - 1.4\beta \left(\frac{0.4}{\epsilon}\right)^{1.26}\,,
		\end{align}
	    where $\Gamma_{0}$ is the torque normalisation at the planet semi-major axis, $a$, which is given by 
	    \begin{align}
	    	\label{eq:gamma0}
	    	&\Gamma_{0} = \left(\frac{q}{h}\right)^2 \Sigma(a) a^4 \Omega^2 \,.
	    \end{align}
		The radial slopes of the density and temperature stratifications are denoted by
	    \begin{align}
	    	&\alpha_{\rm \Sigma}=-\frac{\partial \log \Sigma}{\partial \log r} \quad \mbox{and} \quad \beta=-\frac{\partial \log T}{\partial \log r}\,,
	    \end{align}
	     with $T$ being the temperature.
		The eccentricity correction factors for the torque are calculated using
	    \begin{align}
	    	\label{eq:ecctorquefactors}
	    	&\Delta_{\rm L} = \frac{1-(\frac{e}{2.02\,h})^4}{1+(\frac{e}{2.25\,h})^{1.2}+(\frac{e}{2.84\,h})^6}\,,\\
	    	&\Delta_{\rm C} = \exp \left(\frac{e}{0.5\,h+0.01}\right)\,.
	    \end{align}
	    The calculated torque along with the eccentricity and semi-major axis of the planets give the timescale of angular momentum change
\begin{equation}
   \tau_{\rm L}=L/\Gamma  \,.
\end{equation}
	    The second ingredient, the eccentricity damping timescale $\tau_{\rm e}$, is obtained from Eq.~(11) in \cite{2008A&A...482..677C} and reads
	    \begin{equation}
	    	\label{eq:ecctimescale}
	    	\tau_{\rm e} = 1.282 \left[ 1- 0.14 \left(\frac{e}{h}\right)^2+0.06 \left (\frac{e}{h} \right)^3\right] 
               \, \frac{h^2 a^2 \Omega}{\Gamma_{0}}\,.
	    \end{equation}
	    Finally, we the used accelerations given by \cite{2000MNRAS.315..823P} as
		\begin{equation}
			\label{eq:motion}
			\vec{a} = -\frac{\vec{v}}{\tau_{\rm L}} - 2 \, \frac{(\vec{r} \cdot \vec{v})}{r^2 \tau_{\rm e}} \, \vec{r} \, .
		\end{equation}
		The connection of the actual migration timescale, $\tau_a$, and the angular momentum change, $\tau_L$, is given by
		\begin{equation}
			   \frac{1}{\tau_a} = \frac{2}{\tau_L} + \frac{e^2}{1-e^2} \frac{2}{\tau_e} \,,
		\end{equation}
		see Eq.~(\ref{eq:tortaue}). Only for a circular orbit does the migration timescale equal half of the angular momentum loss timescale. 

	\section{Supplementary figures}
	\label{appendix:B}
	
	In order to keep the main text more concise, we moved supplementary figures to this appendix. We refer to the main text and Table~\ref{tab:models} for details of the specific model setup.

	\begin{figure}
		\centerline{\includegraphics[width=\columnwidth]{ 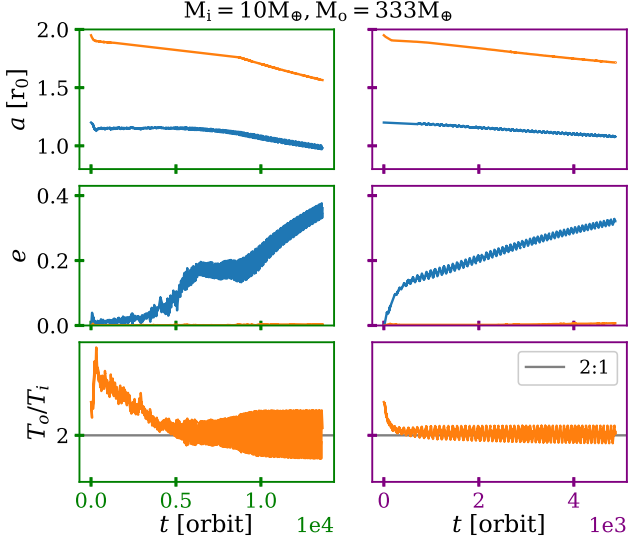}}
		\caption{Similar to Fig.~\ref{fig:21010}, but for the 2pJup models.}
		\label{fig:2101Mj}
	\end{figure}
	
	\begin{figure}
		\centerline{\includegraphics[width=0.9\columnwidth]{ 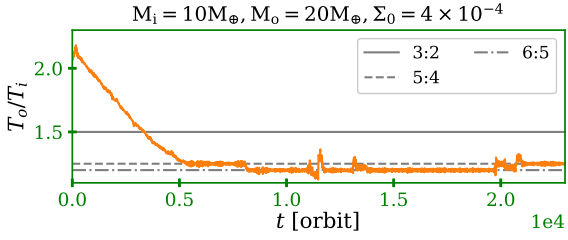}}
		\caption{Orbital period ratio for the 2p20HS \dz~ model, whose surface density is twice that of 2p20.}
		\label{fig:21020HS}
	\end{figure}

	\begin{figure}
		\centerline{\includegraphics[width=\columnwidth]{ 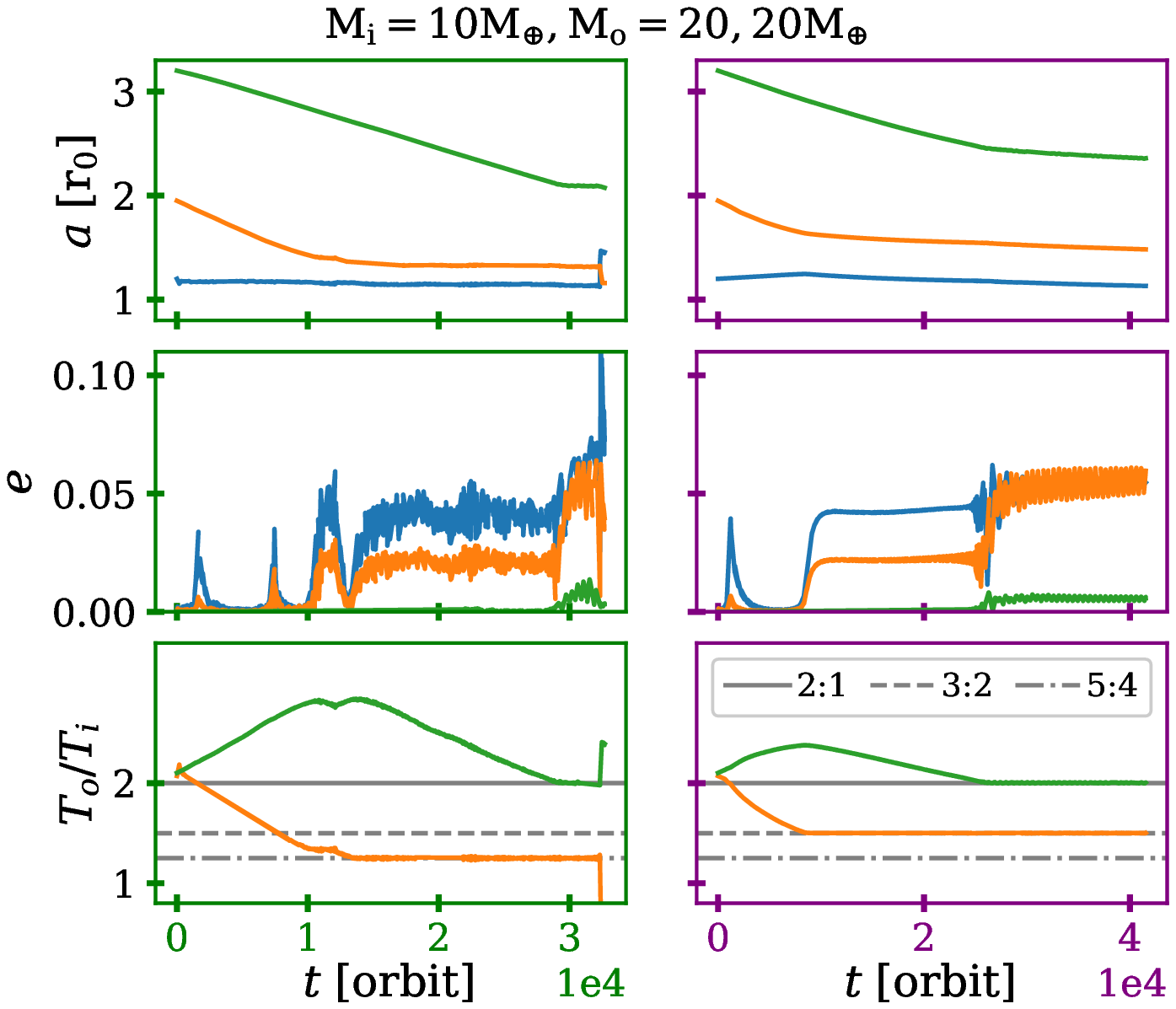}}
		\caption{Similar to Fig.~\ref{fig:3101020}, but for the 3p2020 model.}
		\label{fig:3102020}
	\end{figure}
	
	\begin{figure}
		\centerline{\includegraphics[width=\columnwidth]{ 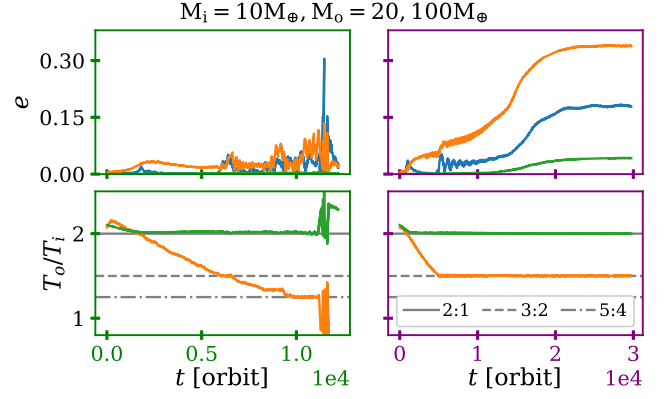}}
		\caption{Eccentricity and orbital period ratio of for the 3p20100 models.}
		\label{fig:31020100}
	\end{figure}
	
	\begin{figure}
		\centerline{\includegraphics[width=\columnwidth]{ 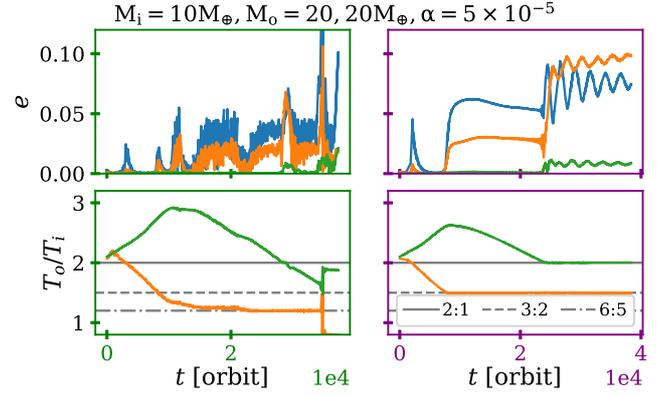}}
		\caption{Evolution of eccentricity and orbital period ratio for 3p2020LA, which has a lower viscosity.}
		\label{fig:3102020LA}
	\end{figure}
	
	\begin{figure}
		\centerline{\includegraphics[width=\columnwidth]{ 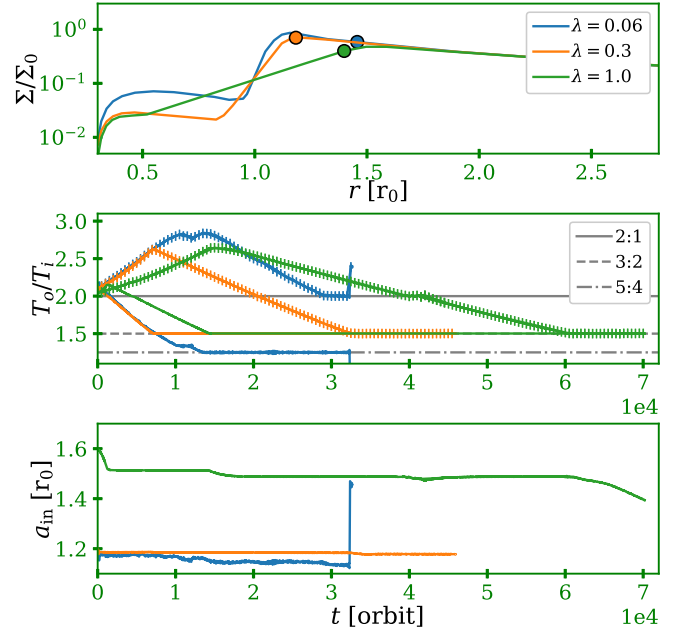}}
		\caption{Similar to Fig.~\ref{fig:cavwidthtratio} (top panels), but for the 3p2020 model. In the middle panel, the lines without and with markers correspond to the results for the inner and outer planet pair. The bottom panel shows the migration of the innermost planet in these three simulations.}
		\label{fig:cavwidth3ptratio}
	\end{figure}

	\begin{figure}
		\centerline{\includegraphics[width=\columnwidth]{ 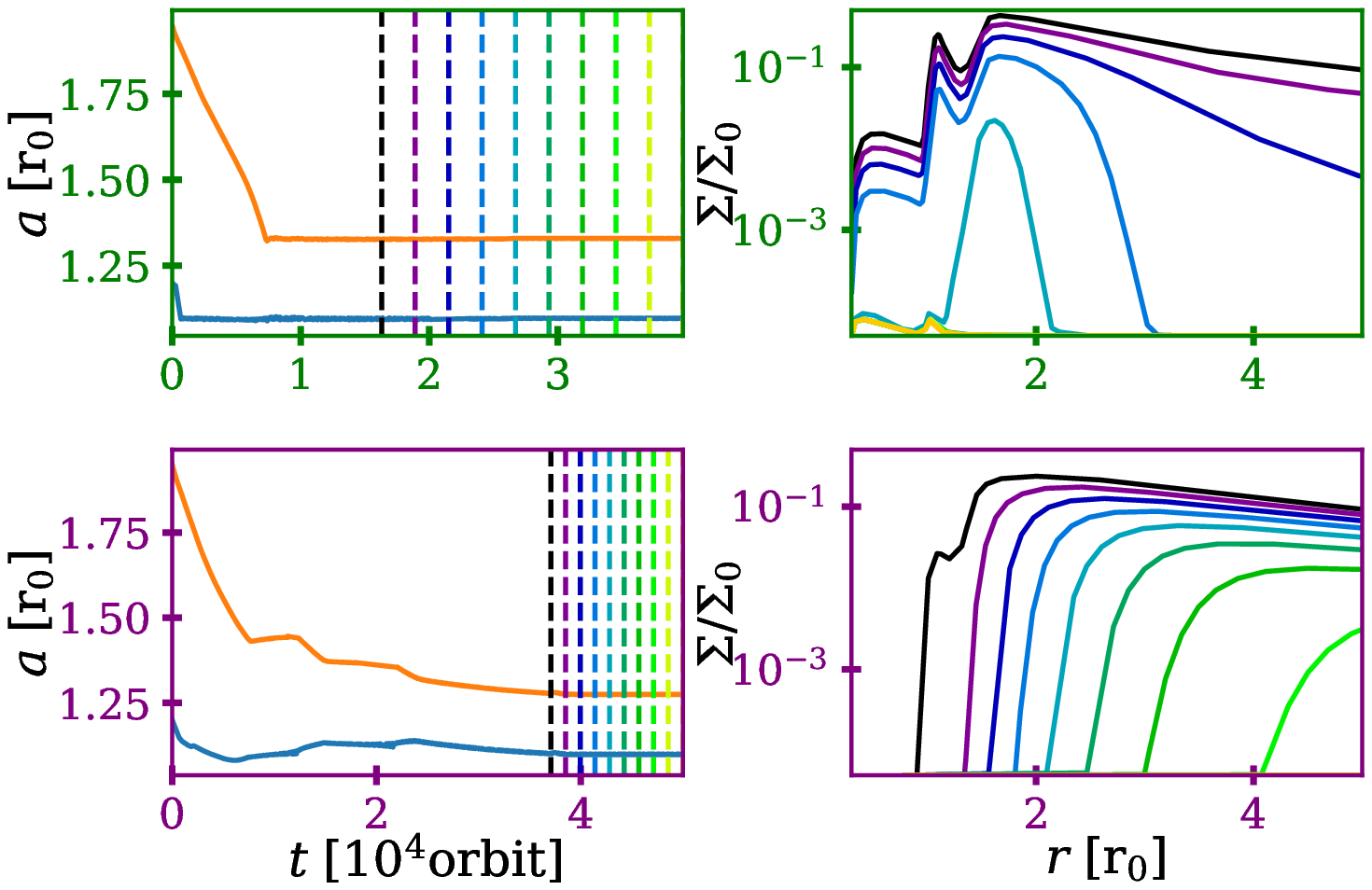}}
		\caption{Similar to Fig.~\ref{fig:photoevap}, but for the 2p20LH models.}
		\label{fig:photoevap03}
	\end{figure}
	
	\end{appendix}
	
\end{document}